\documentclass{emulateapj}

\usepackage{times}
\usepackage{amssymb}
\usepackage{amsmath}
\usepackage{pifont}
\usepackage{graphics}
\usepackage{xcolor}
\usepackage{epsfig}
\usepackage{xspace}

\newcommand{\bjdtdb}{\ensuremath{\rm {BJD_{TDB}}}}
\newcommand{\feh}{\ensuremath{\left[{\rm Fe}/{\rm H}\right]}\xspace}
\newcommand{\teff}{\ensuremath{T_{\rm eff}}\xspace}
\newcommand{\logg}{\ensuremath{\log g_*}\xspace}
\newcommand{\vsini}{\ensuremath{v \sin I_*}\xspace}

\newcommand{\msun}{\ensuremath{\,M_\Sun}\xspace}
\newcommand{\rsun}{\ensuremath{\,R_\Sun}\xspace}
\newcommand{\lsun}{\ensuremath{\,L_\Sun}\xspace}
\newcommand{\mj}{\ensuremath{\,M_{\rm J}}\xspace}
\newcommand{\rj}{\ensuremath{\,R_{\rm J}}\xspace}
\newcommand{\fave}{\langle F \rangle\xspace}
\newcommand{\fluxcgs}{10$^9$ erg s$^{-1}$ cm$^{-2}$\xspace}

\newcommand{\kms}{\,km\,s$^{-1}$\xspace}

\begin{document}
\title{KELT-17\MakeLowercase{b}: A hot-Jupiter transiting an A-star in a misaligned orbit detected with Doppler tomography}

\author{
George Zhou$^{1}$,
Joseph E. Rodriguez$^{2}$, 
Karen A. Collins$^{2,3}$,
Thomas Beatty$^{4,5}$, 
Thomas Oberst$^{6}$,
Tyler M. Heintz$^{6}$,
Keivan G. Stassun$^{2,3}$,
David W. Latham$^{1}$,
Rudolf B. Kuhn$^{7}$,
Allyson Bieryla$^{1}$,
Michael B. Lund$^{2}$, 
Jonathan Labadie-Bartz$^{8}$,
Robert J. Siverd$^{9}$,
Daniel J. Stevens$^{10}$,
B. Scott Gaudi$^{10}$,
Joshua Pepper$^{8}$,
Lars A. Buchhave$^{11}$,
Jason Eastman$^{1}$,
Knicole Col\'{o}n$^{12,13}$, 
Phillip Cargile$^{1}$,
David James$^{14}$,
Joao Gregorio$^{15}$,
Phillip A. Reed$^{16}$,
Eric L. N. Jensen$^{17}$,
David H. Cohen$^{18}$,
Kim K. McLeod$^{18}$,
T.G. Tan$^{19}$,
Roberto Zambelli$^{20}$,
Daniel Bayliss$^{21}$,
Joao Bento$^{22}$,
Gilbert A. Esquerdo$^{1}$,
Perry Berlind$^{1}$,
Michael L. Calkins$^{1}$,
Kirsten Blancato$^{18}$, 
Mark Manner$^{2}$,
Camile Samulski$^{18}$,
Christopher Stockdale$^{23}$,
Peter Nelson$^{24}$,
Denise Stephens$^{25}$,
Ivan Curtis$^{26}$,
John Kielkopf$^{27}$,
Benjamin J. Fulton$^{28,31}$,
D.L. DePoy$^{29}$,
Jennifer L. Marshall$^{29}$,
Richard Pogge$^{10}$,
Andy Gould$^{10}$,
Mark Trueblood$^{30}$, and
Pat Trueblood$^{30}$
}

\affil{$^{1}$Harvard-Smithsonian Center for Astrophysics, 60 Garden St, Cambridge, MA 02138, USA; email: george.zhou@cfa.harvard.edu}
\affil{$^{2}$Department of Physics and Astronomy, Vanderbilt University, 6301 Stevenson Center, Nashville, TN 37235, USA}
\affil{$^{3}$Department of Physics, Fisk University, 1000 17th Avenue North, Nashville, TN 37208, USA}
\affil{$^{4}$Department of Astronomy \& Astrophysics, The Pennsylvania State University, 525 Davey Lab, University Park, PA 16802, USA}
\affil{$^{5}$Center for Exoplanets and Habitable Worlds, The Pennsylvania State University, 525 Davey Lab, University Park, PA 16802, USA}
\affil{$^{6}$Department of Physics, Westminster College, New Wilmington, PA 16172, USA}
\affil{$^{7}$South African Astronomical Observatory, PO Box 9, Observatory 7935, South Africa}

\affil{$^{8}$Department of Physics, Lehigh University, 16 Memorial Drive East, Bethlehem, PA 18015, USA}
\affil{$^{9}$Las Cumbres Observatory Global Telescope Network, 6740 Cortona Dr., Suite 102, Santa Barbara, CA 93117, USA}
\affil{$^{10}$Department of Astronomy, The Ohio State University, Columbus, OH 43210, USA}
\affil{$^{11}$Centre for Star and Planet Formation, Natural History Museum of Denmark, University of Copenhagen, DK-1350 Copenhagen, Denmark}
\affil{$^{12}$NASA Ames Research Center, M/S 244-30, Moffett Field, CA 94035, USA}
\affil{$^{13}$Bay Area Environmental Research Institute, 625 2nd St. Ste 209 Petaluma, CA 94952, USA}
\affil{$^{14}$Cerro Tololo InterAmerican Observatory, Casilla 603, La Serena, Chile}
\affil{$^{15}$Atalaia Group and Crow-Observatory, Portalegre, Portugal}
\affil{$^{16}$Department of Physical Sciences, Kutztown University, Kutztown, PA 19530, USA}
\affil{$^{17}$Department of Physics and Astronomy, Swarthmore College, Swarthmore, PA 19081, USA}
\affil{$^{18}$Wellesley College, Wellesley, MA 02481, USA}
\affil{$^{19}$Perth Exoplanet Survey Telescope, Perth, Australia}
\affil{$^{20}$Societ Astronomica Lunae, Castelnuovo Magra I-19030, Via Montefrancio, 77, Italy}
\affil{$^{21}$Observatoire Astronomique de l'Universit\'e de Gen\`eve, 51 ch. des Maillettes, 1290 Versoix, Switzerland}
\affil{$^{22}$Research School of Astronomy and Astrophysics, Australian National University, Canberra, ACT 2611, Australia}
\affil{$^{23}$Hazelwood Observatory, Victoria, Australia}
\affil{$^{24}$Ellinbank Observatory, Victoria, Australia}
\affil{$^{25}$BYU Department of Physics and Astronomy, N486 ESC, Provo, UT 84602}
\affil{$^{26}$ICO, Adelaide, Australia}
\affil{$^{27}$Department of Physics \& Astronomy, University of Louisville, Louisville, KY 40292, USA}
\affil{$^{28}$Institute for Astronomy, University of Hawai'i at Manoa, HI 96822, USA}
\affil{$^{29}$George P. and Cynthia Woods Mitchell Institute for Fundamental Physics and Astronomy, and Department of Physics and Astronomy, Texas A\&M University, College Station, TX 77843-4242}
\affil{$^{30}$Winer Observatory, Sonoita, AZ 85637, USA}
\affil{$^{31}$NSF Graduate Research Fellow}

\shorttitle{KELT-17b}

\begin{abstract}

We present the discovery of a hot-Jupiter transiting the $V=9.23$\,mag main-sequence A-star KELT-17 (BD+14 1881). KELT-17b is a $1.31_{-0.29}^{+0.28}\,M_\mathrm{J}$, $1.645_{-0.055}^{+0.060}\,R_\mathrm{J}$ hot-Jupiter in a 3.08 day period orbit misaligned at $-115.9\pm4.1$ deg to the rotation axis of the star. The planet is confirmed via both the detection of the radial velocity orbit, and the Doppler tomographic detection of the shadow of the planet during two transits. The nature of the spin-orbit misaligned transit geometry allows us to place a constraint on the level of differential rotation in the host star; we find that KELT-17 is consistent with both rigid-body rotation and solar differential rotation rates ($\alpha < 0.30$ at $2\sigma$ significance). KELT-17 is only the fourth A-star with a confirmed transiting planet, and with a mass of $1.635_{-0.061}^{+0.066}\,M_\odot$, effective temperature of $7454\pm49$\,K, and projected rotational velocity $v\sin I_* = 44.2_{-1.3}^{+1.5}\,\mathrm{km\,s}^{-1}$; it is amongst the most massive, hottest, and most rapidly rotating of known planet hosts. 

\end{abstract}

\maketitle
\section{Introduction}

The properties of planets orbiting high mass stars provide an important piece of the planet formation puzzle. The occurrence rate of giant planets increases with stellar mass \citep[e.g.][]{Johnson:2007,Johnson:2010,Crepp:2011,Jones:2014,Jones:2016}, at least until $\sim 2\, M_\odot$ \citep{Reffert:2015}. Observations of protoplanetary disks also show a correlation between the mass of the host star and the surface density and mass of the protoplanetary disk \citep[e.g.][]{Muzerolle:2003,Natta:2006}, as well as the disk accretion rate \citep{Manara:2016}. As such, the conditions around young, massive stars are more conducive to the formation of giant planet embryos \citep[e.g.][]{Liu:2016}, and may even lead to more massive planets being formed \citep{Mordasini:2012}. 

Despite the apparent ease of giant planet formation around massive stars, only three transiting planets have been confirmed around A stars to date: WASP-33b \citep{Collier:2010b,Johnson:2015}, KOI-13b \citep{Szabo:2011,Shporer:2011,Johnson:2014}, and HAT-P-57b \citep{Hartman:2015}. Confirming planets around these stars is difficult via traditional techniques: in addition to the low mass and radius ratios of these systems (therefore low radial velocity amplitudes and transit depths), main sequence A-stars have rapid rotation rates and few metal spectral lines, inhibiting precise radial velocity measurements that are typically necessary for the planet confirmation. One successful approach is to perform radial velocity searches for planets around `retired A-stars' \citep[e.g.][]{Johnson:2010,Wittenmyer:2011} -- giants and sub-giants with masses $>1.6\,M_\odot$ that have spun-down over their post-main-sequence evolution, and allow precise radial velocity measurements to be made. These surveys have revealed some intriguing trends, such as the apparent lack of high eccentricity warm-Jupiters around sub-giants \citep{Jones:2014}.

Transiting planets offer an unique set of opportunities, such as the characterization of planet radius, orbital orientation, and atmospheric properties, that are not available to planets detected by radial velocities only. A sample of well characterized planets around massive stars is necessary to understand the mass-dependence of planet properties. The Kilodegree Extremely Little Telescope (KELT) \citep{Pepper:2007} is designed to target planets orbiting stars with brightnesses of $8 < V_\mathrm{mag} < 10$: systems around bright stars that are conducive to follow-up characterization. As discussed in \citet{Bieryla:2015}, a direct result of this KELT sample selection is that 55\% of KELT-North targets are hotter than 6250\,K, with masses $\gtrsim 1.3\,M_\odot$ and median rotational velocities of $\gtrsim 20$\,km\,s$^{-1}$ \cite[inferred from the Kepler sample of stellar rotational velocities in][]{Nielsen:2013}. A similar stellar sample will also be surveyed by the TESS full frame dataset \citep{Ricker:2014}. Strategies for confirming planets around massive stars from the KELT survey are therefore directly transferable to future planet candidates from TESS. 

Transiting planets around rapidly rotating, high mass stars can be confirmed via Doppler tomography. During a transit, the planet occults parts of the rotating stellar disk, thereby distorting the observed spectral line profile of the star. For relatively slowly rotating host stars, this results in a net shift in the apparent velocity of the host star -- the Rossiter-McLaughlin effect \citep{Rossiter:1924,McLaughlin:1924}. In the cases where the rotational broadening of the star is significantly higher than other broadening factors, the shadow of the planet can be resolved as an intrusion  in the rotationally broadened line profile of the star, yielding a Doppler tomographic detection of the planet. In addition to the three planets around A-stars confirmed via Doppler tomography, detections have been achieved for 9 more planets: WASP-3b \citep{Miller:2010}, WASP-38b \citep{Brown:2012}, CoRoT-11b \citep{Gandolfi:2012}, HAT-P-2b and Kepler-25c \citep{Albrecht:2013}, KOI-13b \citep{Johnson:2014}, KOI-12b \citep{Bourrier:2015}, KELT-7b and HAT-P-56b \citep{Bieryla:2015,Zhou:2016}. The depth and width of the spectroscopic shadow of the planet is directly correlated with the planet-star radius ratio. In the cases where the depth agrees with that from the photometric transit, we can rule out blend scenarios often associated with transiting planet candidates. This is especially useful in eliminating the scenarios of background eclipsing binaries, where a Doppler tomographic observation will yield no planet detection. Subsequent out-of-transit radial velocities, at the km\,s$^{-1}$ level, are then taken to constrain the nature of the orbiting companion.

In this paper, we report the discovery of KELT-17b, a hot-Jupiter transiting a rapidly rotating $(v\sin I_* = 44\,\mathrm{km\,s}^{-1})$ A-star. KELT-17b was discovered in the equatorial field jointly surveyed by the KELT-North \citep{Pepper:2007} and KELT-South \citep{Pepper:2013} observatories. The discovery involves a series of photometric follow-up observations that confirmed and characterized the transit light curve, and spectroscopic monitoring that constrained the mass of the system. Finally, blend false positive scenarios were ruled out by two in-transit spectroscopic observations that confirmed the Doppler tomographic signal induced by the transiting planet.

\section{Discovery and Follow-Up Observations}

\subsection{KELT-South and KELT-North}
\begin{figure*}
    \centering
    \includegraphics[width=15cm]{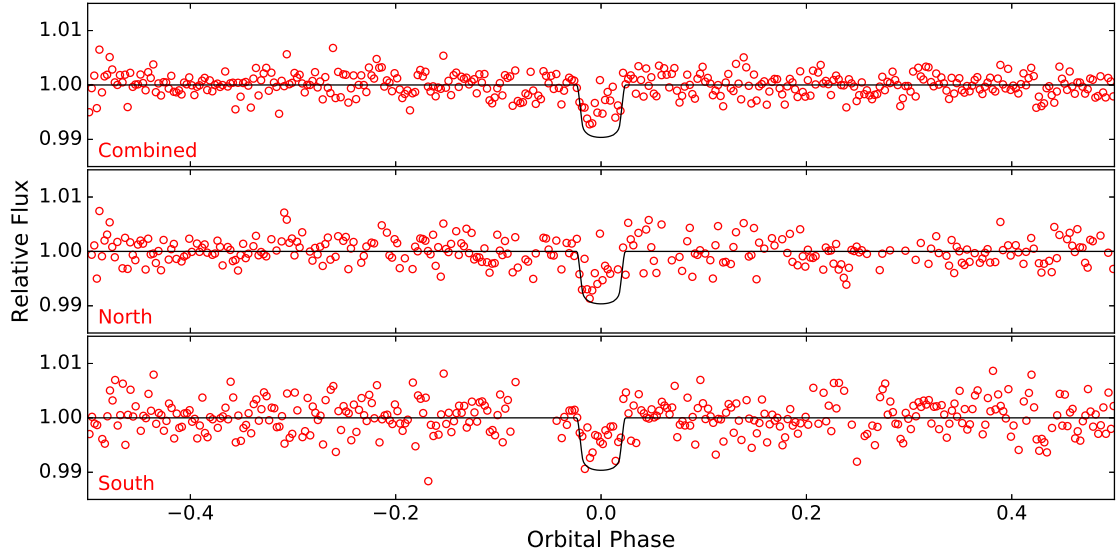}
    \caption{Discovery light curves from the KELT survey. The data points show the discovery light curve, binned at intervals of 0.0025 in phase. The combined light curve from the North and South telescopes are shown on the top panel. Subsequent panels show the light curves from each site. The KELT discovery dataset was not used to constrain the planet parameters in the global fit (Section~\ref{sec:exofast}). The transit depth from the discovery light curves are diluted due to the application of TFA, which acts to dampen any modulation in the light curve. The final transit model from the EXOFAST global analysis, inferred via higher precision follow-up photometry, is plotted in black for reference.}
    \label{fig:KS_discoverylc}
\end{figure*}

KELT-17, the first exoplanet host discovered through the combined observations of both the KELT-North and KELT-South telescopes, is located in KELT-South field 06 (KS06) and KELT-North field 14 (KN14) which are both centered on $\alpha =$ 07$^{h}$ 39$^{m}$ 36$^{s}$ $\delta =$ $+03\degr$ 00$\arcmin$ 00$\arcsec$ (J2000). At the time of identification, the post-processed KELT data set included 2092 images from KN14 taken between UT 2011 October 11 and UT 2013 March 26 and 2636 images from KS06 taken between UT2010 March 02 and 2013 May 10. The image reduction, light curve extraction, and candidate selection processes are described in \citet{Siverd:2012, Kuhn:2016}. In brief, calibrated images are processed to light curves using a heavily modified version of the ISIS image subtraction package \citep{Alard:1998,Alard:2000}. Extracted light curves are outlier-clipped, smoothed with a 90-day median window, and finally detrended with the Trend Filtering Algorithm \citep{Kovacs:2005}. This process was performed independently for the KELT-North and KELT-South data sets. Objects in common between the two fields were identified and given a new KELT-Joint field 06 (KJ06) designation. The candidate selection process was then run on these final combined light curves. KJ06C006046=KELT-17 emerged as a top candidate in the joint analysis of field KJ06. KELT-17 (TYC 807-903-1, 2MASS J08222820+1344071) is located at $\alpha =$ 08$^{h}$ 22$^{m}$ 28$\fs$21 $\delta =$ +13$\degr$ 44$\arcmin$ 07$\arcsec$2 (J2000). A list of the photometric and kinematics parameters for KELT-17 is shown in Table \ref{tbl:Host_Lit_Props}. The box-fitting least squares (BLS) \citep{Kovacs:2002} periodicity algorithm was used to search for candidates in KJ06. Candidates are selected according to statistics produced with the VARTOOLS \citep{Hartman:2016} implementation of BLS, and from statistics calculated as per \citet{Pont:2006} and \citet{Burke:2006}. Table \ref{tbl:BLS_Selection_Criteria} shows the criteria and results for the KELT-17b candidate selection. The discovery light curves from both KELT-North and KELT-South are shown in Figure~\ref{fig:KS_discoverylc}.

\begin{table*}
\centering
\caption{Magnitudes and Kinematics of KELT-17}
\label{tbl:Host_Lit_Props}
\begin{tabular}{llccc}
  \hline
  \hline
\hline
  Parameter & Description & KELT-17 Value & Source & Reference(s) \\
Names& 					&  TYC 807-903-1	& 		&			\\
			& 					& 2MASS J08222820+1344071 		& 		&			\\
			& 					& BD+14 1881 		& 		&			\\
			&					&				&		&			\\
$\alpha_{J2000}$	&Right Ascension (RA)& 08:22:28.21		& Tycho-2	& \citet{Hog:2000}	\\
$\delta_{J2000}$	&Declination (Dec)& +13:44:07.2			& Tycho-2	& \citet{Hog:2000}	\\
			&					&				&		&			\\
FUV & Far UV flux & 16.565 $\pm$ 0.034 & GALEX & \citet{Bianchi:2011}\\
NUV & Near UV flux & 13.261 $\pm$ 0.005 & GALEX & \citet{Bianchi:2011}\\
			&					&				&		&			\\
$u'$			& & 11.027 $\pm$ 0.001		& SDSS	& \citet{Abazajian:2009}	\\
B			& & 9.553 $\pm$ 0.025		& ASCC	& \citet{Kharchenko:2001}	\\
B$_T$			&Tycho B$_T$ magnitude& 9.53 $\pm$ 0.02		& Tycho-2	& \citet{Hog:2000}	\\
V$_T$			&Tycho V$_T$ magnitude& 9.23 $\pm$ 0.02		& Tycho-2	& \citet{Hog:2000}	\\
V &         & 9.286 $\pm$ 0.051 & TASS & \citet{Droege:2006} \\
$r'$			&    & 9.223 		&   CMC15	&   \citet{Evans:2002}	\\
$I_C$ & & 8.948 $\pm$ 0.039 & TASS & \citet{Droege:2006} \\
			&					&				&		&			\\
J			&2MASS magnitude& 8.745 $\pm$ 0.027		& 2MASS 	& \citet{Cutri:2003, Skrutskie:2006}	\\
H			&2MASS magnitude& 8.697 $\pm$ 0.042	& 2MASS 	& \citet{Cutri:2003, Skrutskie:2006}	\\
K			&2MASS magnitude& 8.646 $\pm$ 0.018		& 2MASS 	& \citet{Cutri:2003, Skrutskie:2006}	\\
			&					&				&		&			\\
\textit{WISE1}		&WISE passband& 8.616 $\pm$ 0.023		& WISE 		&\citet{Cutri:2012}	\\
\textit{WISE2}		&WISE passband& 8.644 $\pm$ 0.02		& WISE 		& \citet{Cutri:2012}\\
\textit{WISE3}		&WISE passband& 8.630 $\pm$ 0.053		& WISE 		& \citet{Cutri:2012}	\\
\textit{WISE4}		&WISE passband& 8.678 $\pm$ 0.384		& WISE 		& \citet{Cutri:2012}	\\
			&					&				&		&			\\
$\mu_{\alpha}$		& Proper Motion in RA (mas yr$^{-1}$)	& -22.9 $\pm$ 1.1	& NOMAD		& \citet{Zacharias:2004} \\
$\mu_{\delta}$		& Proper Motion in DEC (mas yr$^{-1}$)	&  -0.7 $\pm$ 1.0	& NOMAD		& \citet{Zacharias:2004} \\
			&					&				&		&			\\
U$^{*}$ & Space motion (\kms) &   -25.6 $\pm$ 0.9 & &  This work \\
V & Space motion (\kms) & 3.3 $\pm$ 0.9 & & This work \\
W & Space motion (\kms) & -0.1 $\pm$ 1.3 &  &  This work \\
Distance & Distance (pc) & 210 $\pm$ 10 &  &  This work \\
RV & Absolute Radial Velocity (\kms) &  28.0 $\pm$ 0.1 & &  This work \\
 \hline
\hline
\hline
\end{tabular}

 \footnotesize \textbf{\textsc{NOTES}}\\

\footnotesize $^{*}$U is positive in the direction of the Galactic Center 
\end{table*}

\begin{table}
 \caption{KELT discovery selection criterion}
\small
\label{tbl:BLS_Selection_Criteria}
 \begin{tabular}{@{}llll}
    \hline
    Statistic  &  Selection  & KELT-17b\\
   &  Criteria & /KJ06C006046\\
    \hline
    Signal detection \dotfill & SDE $>$ 7.0 & 10.56395\\
      efficiency\dotfill & & & \\
  Signal to pink-noise\dotfill & SPN $>$ 7.0 &9.27154\\
    Transit depth\dotfill & $\delta <$ 0.05 & 0.00433\\
    $\chi^2$ ratio\dotfill &  $\displaystyle\frac{\Delta\chi^2}{\Delta\chi^2_{-}} >$ 1.5 & 1.72 \\
    Duty cycle\dotfill & q $<$ 0.1 &0.04\\
    \hline
 \end{tabular}
\end{table}

\subsection{Photometric Follow-up}
\label{sec:Follow-up_Photometry}
\begin{figure}
\vspace{.0in}
\includegraphics[width=1\linewidth,height=5in]{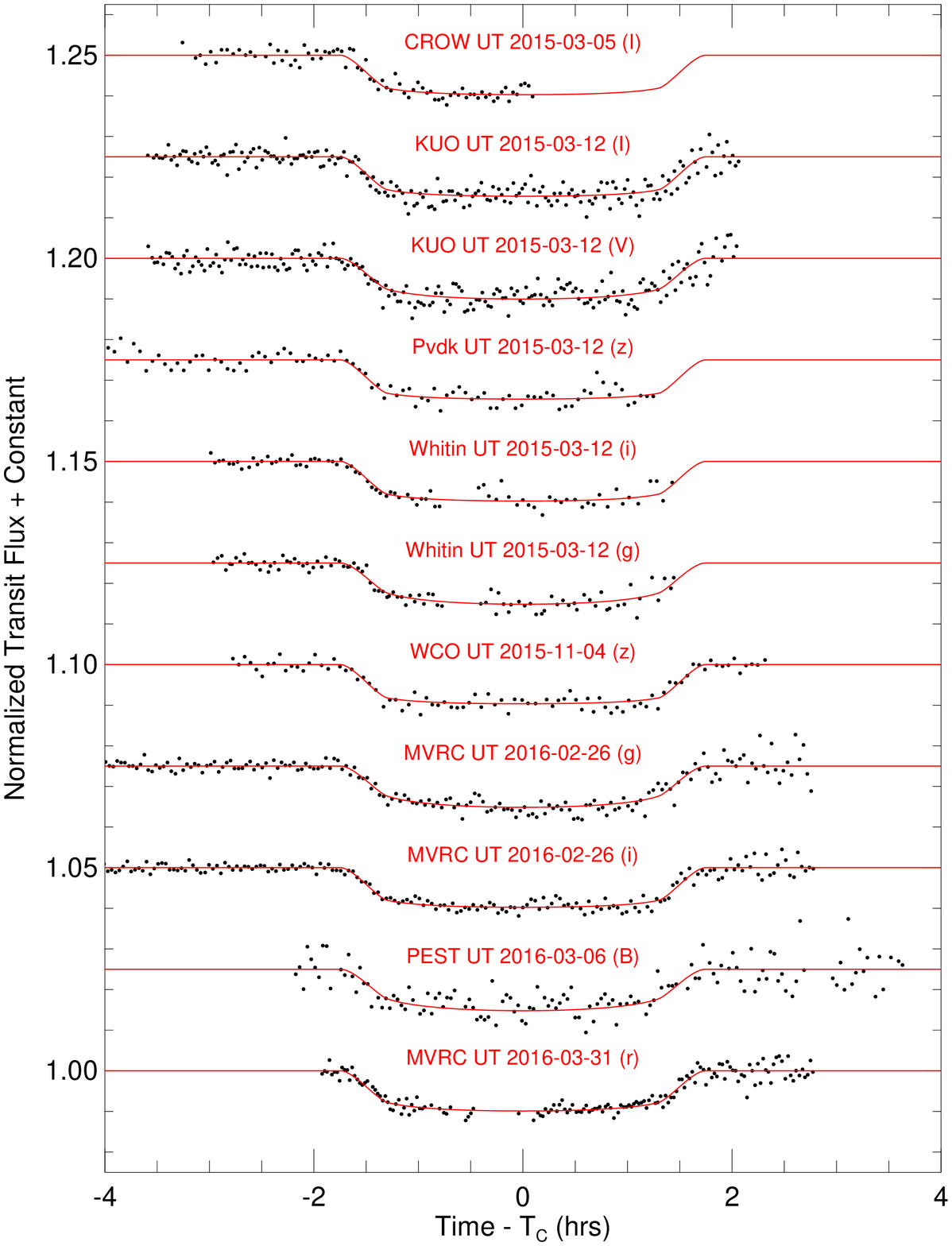}
\vspace{-.25in}

\includegraphics[width=1\linewidth]{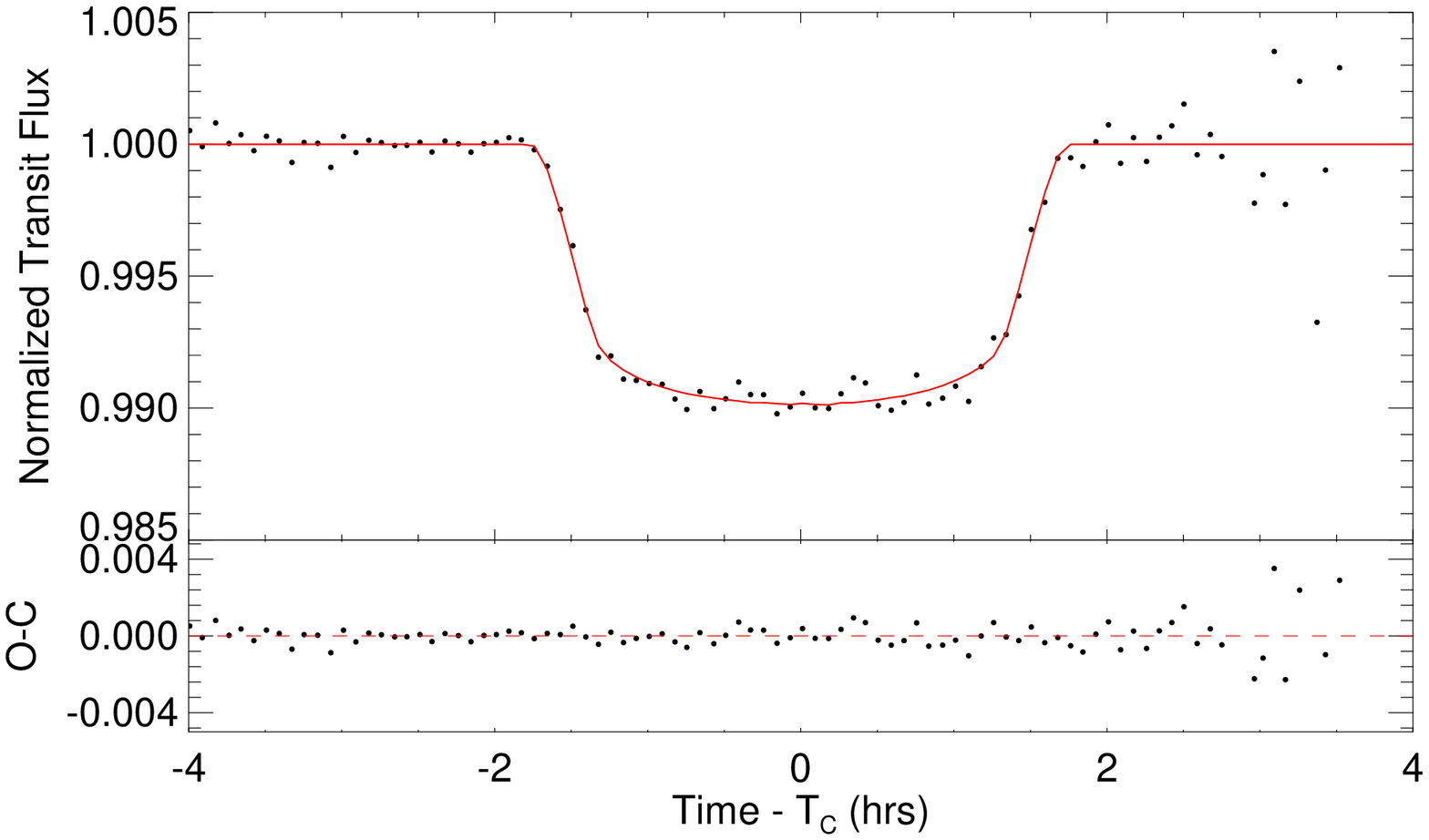}
\caption{(Top) The individual KELT follow-up Network observations of KELT-17b. The best fit model for each light curve is shown in red. (Bottom) All follow-up transits combined and binned in 5 minute intervals to best represent the transit features. These binned data are shown for display only, and were not used in the analysis. The combined and binned models from each transit are represented by the red line.}
\label{fig:All_light curve} 
\end{figure}

To confirm the source, refine the transit depth, duration, period, and eliminate false positive scenarios, we obtained higher spatial resolution and precision  photometric follow-up observations of KELT-17b in multiple filters. These datasets are uniformly reduced using AstroImageJ \citep[AIJ,][]{Collins:2016}. These light curves are presented in Figure~\ref{fig:All_light curve}. A description of each observatory is below. See Table \ref{tbl:detrending_parameters} for a list of the observations and their parameters that were used in the global fit.

\subsubsection{CROW}
An $I$ band transit was observed on UT 2015 March 05 at the Canela's Robotic Observatory (CROW) with the 0.3\,m SCT12 telescope, remotely operated from Portalegre, Portugal. Observations were acquired with the ST10XME CCD camera, with a $30\arcmin$ $\times$ 20$\arcmin$ field of view and a $0\farcs86$ pixel scale.

\subsubsection{Kutztown}
A full multi-color ($V$ and $I$) transit of KELT-17b was observed on UT 2015 March 12 at Kutztown University Observatory (KUO), located on the campus of Kutztown University in Kutztown, Pennsylvania. KUO's main instrument is the 0.6\,m Ritchey-Chr\'{e}tien optical telescope with a focal ratio of $f/8$. The imaging CCD (KAF-6303E) camera has an array of 3K$\times$2K ($9\,\mu$m) pixels and covers a field of view of $19\farcm5 \times 13\farcm0$. 

\subsubsection{Swarthmore}
The Peter van de Kamp Observatory (PvdK) at Swarthmore College (near Philadelphia) houses a 0.62-m Ritchey-Chretien reflector with a 4K$\times$4K pixel Apogee CCD. The telescope and camera together have a 26$\arcmin$ $\times$ 26$\arcmin$ field of view and a $0\farcs61$ pixel scale. PvdK observed KELT-17b on UT 2015 March 12 in the SDSS $z^{\prime}$ filter.

\subsubsection{Whitin}
KELT-17b was observed in both $g^{\prime}$ and $i^{\prime}$ on UT 2015 March 12 at Wellesley College's Whitin Observatory in Massachusetts.  The telescope is a 0.6 m Boller and Chivens with a DFM focal reducer yielding an effective focal ratio of f/9.6. We used an Apogee U230 2K$\times$2K camera with a $0\farcs58$\,pixel$^{-1}$ scale and a $20\arcmin \times 20\arcmin$ field of view.

\subsubsection{WCO}
One full transit of KELT-17b was observed from the Westminster College Observatory (WCO), PA, on UT 2015 November 4 in the $z'$ filter. The observations employed a 0.35\,m f/11 Celestron C14 Schmidt-Cassegrain telescope and SBIG STL-6303E CCD with a $\sim$ 3K$\times$2K array of 9 $\mu$m $\,$ pixels, yielding a $24\arcmin \times 16\arcmin$ field of view and $1\farcs4\, \mathrm{pixel}^{-1}$ image scale at $3\times3$ pixel binning. The stellar FWHM was seeing-limited with a typical value of $\sim 3.2\arcsec$. 

\subsubsection{MVRC}

Three full transits of KELT-17b were observed on UT 2016 February 26 ($g^{\prime}$ and $i^{\prime}$) and UT 2016 March 31 ($r^{\prime}$) using the Manner-Vanderbilt Ritchie-Chr\'{e}tien (MVRC) telescope located at the Mt. Lemmon summit of Steward Observatory, AZ. The observations employed a 0.6 m f/8 RC Optical Systems Ritchie-Chr\'{e}tien telescope and SBIG STX-16803 CCD with a 4K$\times$4K array of 9 $\mu$m pixels, yielding a $26\arcmin \times 26\arcmin$ field of view and $0\farcs39$ pixel$^{-1}$ image scale. The telescope was heavily defocused for all three observations resulting in a typical stellar FWHM of $\sim$ 17$\arcsec$.

\subsubsection{PEST}
The PEST (Perth Exoplanet Survey Telescope) observatory is a backyard observatory owned and operated by Thiam–Guan (TG) Tan, located in Perth, Australia. It is equipped with a 0.3\,m Meade LX200 SCT $f/10$ telescope with focal reducer yielding $f/5$ and a SBIG ST-8XME CCD camera. The telescope and camera combine to have a 31$\arcmin$ $\times$ 21$\arcmin$ field of view and a $1\farcs2$ pixel scale. PEST observed KELT-17b on UT 2016 March 06 in the $B$ band.


\begin{table*}
 \centering
 \caption{Photometric follow-up observations and the detrending parameters found by AIJ for the KELT-17 global fit.}
 \label{tbl:detrending_parameters}
 \begin{tabular}{lllllllll}
    \hline
    \hline
 Observatory & Date (UT) & Filter & FOV & Pixel Scale & Exposure (s)  & Detrending parameters for global fit \\
    \hline
CROW & 2015 March 05 & $I$ & $30\arcmin$ $\times$ 20$\arcmin$ & $0\farcs86$ & 90 & Meridian Flip, Airmass, BJD \\
KUO &  2015 March 12 & $I$ & $19\farcm5\times13\arcmin$ & $0\farcs76$ & 30 & Airmass, BJD\\
KUO &  2015 March 12 & $V$ & $19\farcm5 \times 13\arcmin$ & $0\farcs76$ & 30 & BJD, X Centroid, Sky Background \\
Pvdk &  2015 March 12 & $z^{\prime}$ & 26$\arcmin$ $\times$ 26$\arcmin$ & $0\farcs61$ & 60 & Airmass\\
Whitin &  2015 March 12 & $g^{\prime}$ & $20\arcmin$ $\times$ 20$\arcmin$ & $0\farcs58$ & 48-80 & FWHM, Total Counts, PSF Roundness \\
Whitin &  2015 March 12 & $i^{\prime}$ & $20\arcmin$ $\times$ 20$\arcmin$ & $0\farcs58$ &60-100& BJD, FWHM, Sky Background, Airmass, Total Counts\\
WCO  & 2015 November 04 &  $z^{\prime}$ & $24\arcmin$ $\times$ 16$\arcmin$ & $1\farcs4$ &180 &  BJD, Airmass, Total Counts\\
MVRC & 2016 February 26 & $g^{\prime}$ & $26\farcm8 \times 26\farcm8$ & $0\farcs39$ & 30 & Airmass, FWHM, X Centroid, Y Centroid\\
MVRC & 2016 February 26 & $i^{\prime}$ & $26\farcm8 \times 26\farcm8$ & $0\farcs39$ & 60 & X Centroid, Y Centroid\\
PEST & 2016 March 06 & $B$ & 31$\arcmin$ $\times$ 21$\arcmin$&  $1\farcs2$ & 120 & Target raw counts, FWHM, X Centroid\\
MVRC & UT 2016 March 31 & $r^{\prime}$ & $26\farcm8 \times 26\farcm8$ & $0\farcs39$ & 30 & Airmass, Total Counts, Sky Background\\
     \hline
    \hline
 \end{tabular}
\begin{flushleft}
  \footnotesize \textbf{\textsc{NOTES}} \\
  \footnotesize All the follow-up photometry presented in this paper is available in machine-readable form in the online journal.
  \end{flushleft}
\end{table*}

\subsection{Spectroscopic Follow-up}
\label{sec:spec_fu}
\begin{table*}
 \centering
 \caption{Spectroscopic follow-up observations}
 \label{tbl:spectroscopic_parameters}
 \begin{tabular}{llllllll}
    \hline
    \hline
Telescope/Instrument & Date Range & Type of Observation & Resolution & Wavelength Range ($\AA$) & Mean S/N per res. element & Epochs\\
    \hline
ANU 2.3m/WiFeS & 02/2015 & Low Resolution Spectral Typing & 3000 & 3500--6000 & 135 & 1\\
ANU 2.3m/WiFeS & 02/2015 & Low Resolution Radial Velocity & 7000 & 5200--7000 & 100 & 3\\
FWO 1.5m/TRES & 04/2015 -- 04/2016 &  High Resolution &$\approx$44,000 & 3900--9100 & $\sim$100 & 74 \\
  \hline
    \hline
 \end{tabular}
\end{table*}  

A series of spectroscopic follow-up observations were performed to characterize the KELT-17 system, they are summarized in Table~\ref{tbl:spectroscopic_parameters}.

In order to search for signs of stellar-mass companions of KELT-17, we performed low resolution, high signal-to-noise reconnaissance spectroscopic follow-up of KELT-17 using the Wide Field Spectrograph \citep[WiFeS,][]{Dopita:2007} on the ANU 2.3\,m telescope at Siding Spring Observatory, Australia in February 2015. WiFeS is an integral field spectrograph employing $1\arcsec$ width slitlets to provide a field of view of $12''\times 38''$ when read out in the `stellar' mode. Follow-up with WiFeS allowed us to obtain multi-epoch spectra for the target and all nearby stars within one KELT pixel, helping to eliminate astrophysical blend scenarios, such as nearby eclipsing binaries, that may mimic the signal of a transiting hot-Jupiter \citep{Bayliss:2013}. Stellar classification using the flux calibrated WiFeS spectrum provided an initial estimate for the stellar parameters of $T_\mathrm{eff} = 7200\pm200$\,K, $\log g_* = 4.0\pm0.4$ dex, $\mathrm{[Fe/H]} = -0.5 \pm 0.4$ dex. Three additional multi-epoch observations constrained the radial velocity variation of the target to be $<2\,\mathrm{km\,s}^{-1}$, indicating any orbiting companion responsible for the transit must be of substellar mass. 

Following candidate vetting with WiFeS, in-depth spectroscopic characterization of KELT-17 was performed by the Tillinghast Reflector Echelle Spectrograph (TRES) on the 1.5\,m telescope at the Fred Lawrence Whipple Observatory, Mount Hopkins, Arizona, USA. TRES has a wavelength coverage of 3900--9100\,\AA\, over 51 echelle orders, with a resolving power of $\lambda/\Delta \lambda \equiv R = 44000$. A total of 12 out-of-transit observations were taken to characterize the radial velocity orbital variations exhibited by KELT-17. The wavelength solutions are derived from Th-Ar hollow cathode lamp exposures that bracket each object spectrum. Each observation consists of three consecutive exposures, totalling $\sim 540$\,s in exposure time, combined to enable cosmic-ray removal. 

\begin{figure}
\includegraphics[width=1\linewidth]{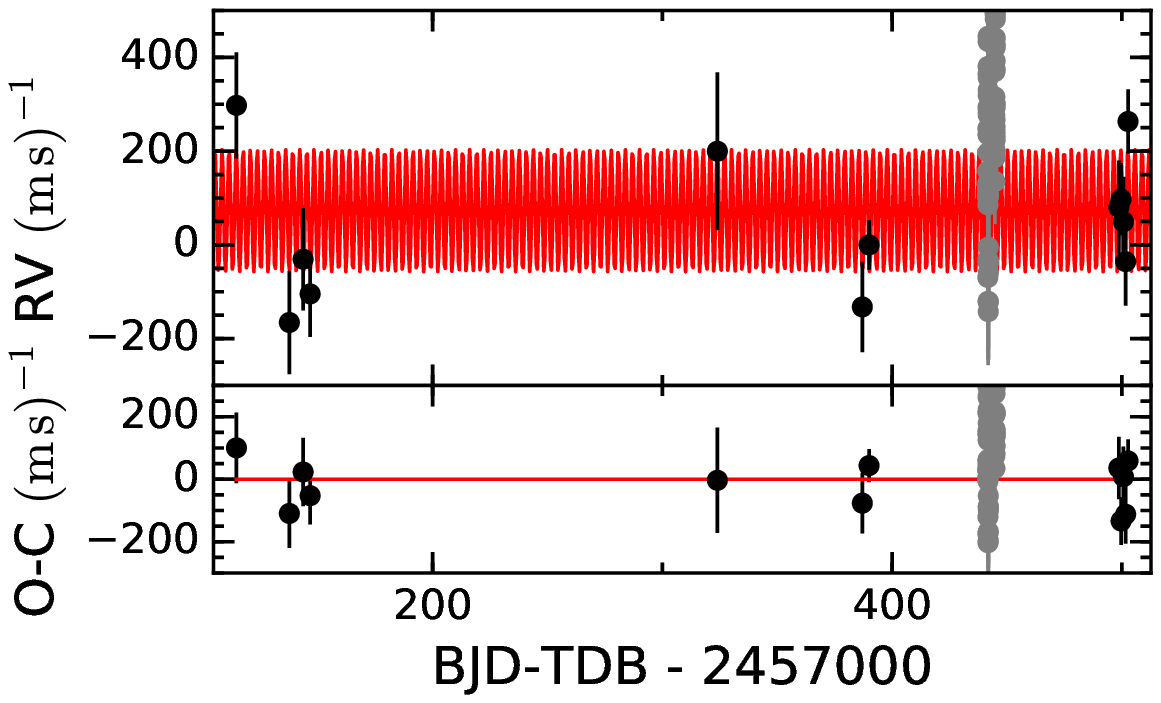}
 \vspace{-.2in}

\includegraphics[width=1\linewidth]{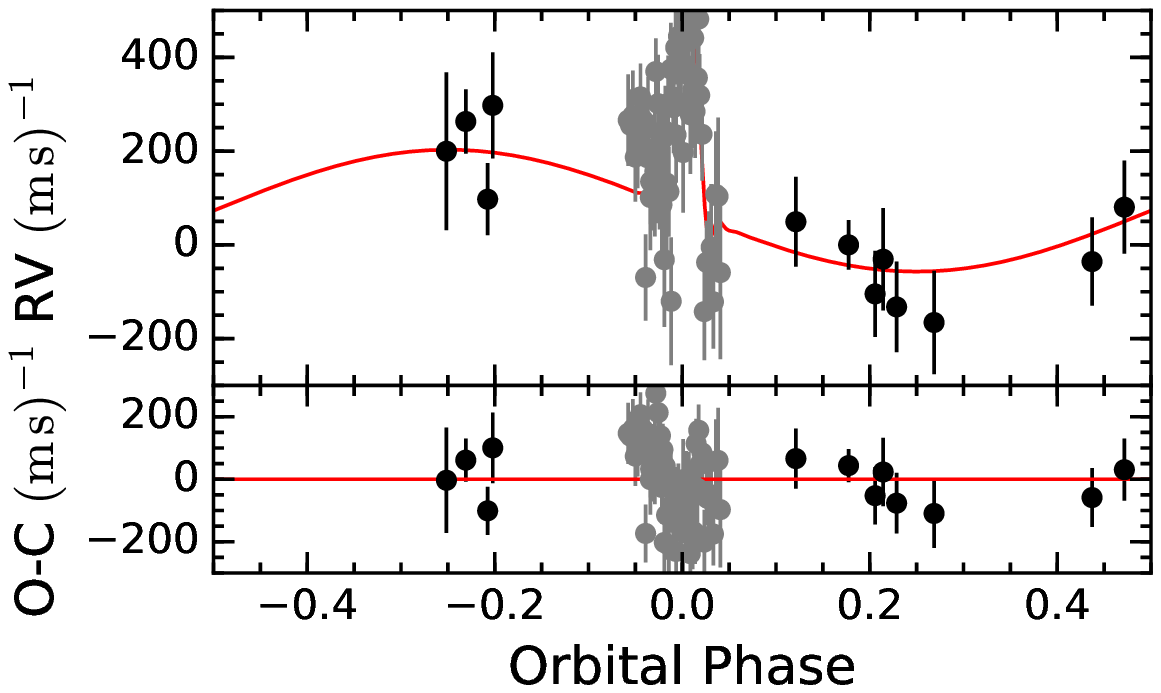}
 \vspace{-.2in}

\includegraphics[width=1\linewidth]{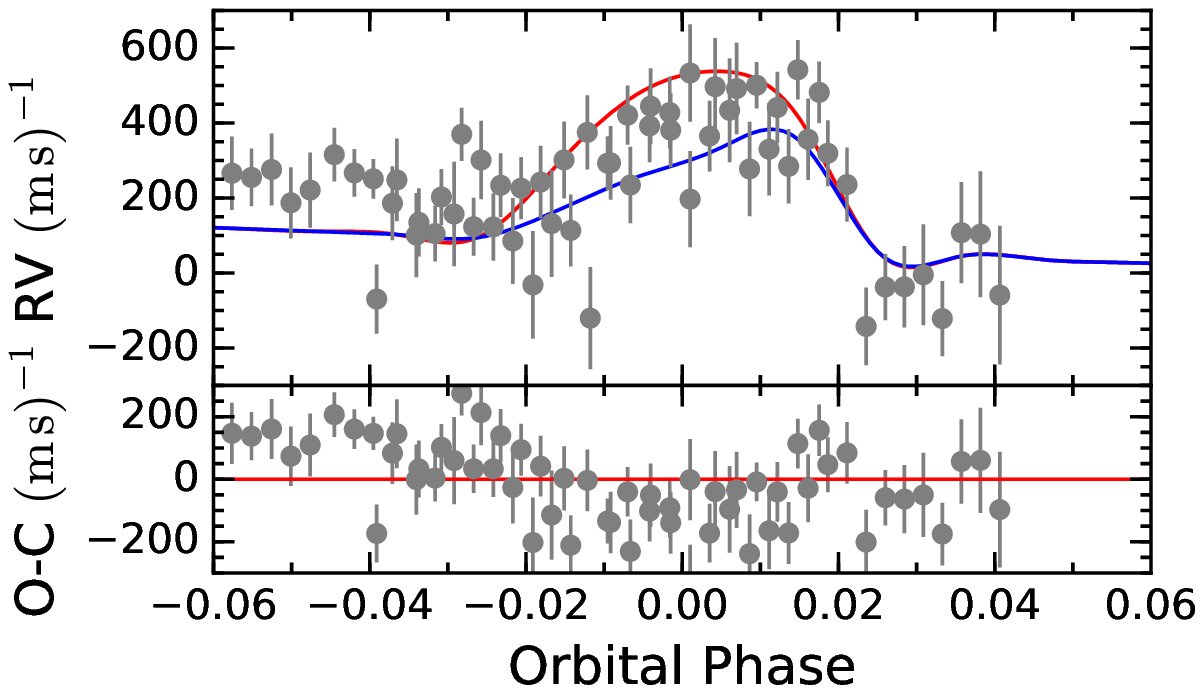}
 \vspace{-.2in}

\caption{TRES radial velocities of KELT-17. The TRES out-of-transit points are shown in black, in-transit points in grey. Only the out-of-transit radial velocities are used in the global fit. In-transit Doppler tomographic analysis, rather than radial velocities, were used to determine the spin-orbit alignment parameters (Section~\ref{sec:exofast}).  (Top) Radial velocities shown as a function of time. (Middle) Velocities shown as a function of orbital phase. (Bottom) Zoomed in view of the in-transit velocities, with the predicted Rossiter-McLaughlin model \citep{Hirano:2011}, with parameters determined from Doppler tomography, over-plotted. Note the in-transit velocities show a clear Rossiter-McLaughlin signal, and are largely consistent with the predicted model. Potential differences may be due to approximations in the Rossiter-McLaughlin model used to estimate the in-transit velocities.}
\label{fig:K11RV} 
\end{figure}

The process of spectral extraction, reduction, and radial velocity analyses are similar to those described in \citet{Buchave:2010}. Absolute radial velocities are obtained by cross correlating the Mg b line region against a synthetic template spectrum generated using the \citet{Kurucz:1992} atmosphere models. The Mg b velocities are used to determine the absolute velocity offsets presented in Tables~\ref{tbl:Host_Lit_Props} and \ref{tbl:KELT-17b}. Precise relative radial velocities are obtained by cross correlating multiple echelle orders of each spectrum to synthetic spectral templates, and weight-averaging the resulting velocities. We adopt the relative radial velocities from the multi-order cross correlations for our radial velocity orbit analysis. The radial velocity orbit measured by TRES is shown in Figure~\ref{fig:K11RV}, and individual radial velocity measurements are also presented in Table~\ref{tab:RVs}.

\begin{table*}
{\tiny
    \caption{Relative radial velocities for KELT-17}
    \label{tab:RVs}
    \centering
    \begin{tabular}{rrrrr}
    \hline\hline
BJD (UTC) & Relative RV (m\,s$^{-1}$) & RV Error (m\,s$^{-1}$) & Exp. Time (s) & SNRe$^a$\\ 
\hline
2457114.63339 & 298 & 113 & 300 & 77.8\\ 
2457137.64461 & -166 & 110 & 360 & 77.0\\ 
2457143.63772 & -31 & 109 & 360 & 72.3\\ 
2457146.69136 & -105 & 92 & 360 & 62.7\\ 
2457323.93301 & 200 & 168 & 480 & 45.9\\ 
2457387.01541 & -132 & 97 & 360 & 84.5\\ 
2457389.93777 & 0$^b$ & 53 & 660 & 136.0\\ 
$^\star$2457441.63378 & -188 & 92 & 540 & 84.4\\ 
$^\star$2457441.64173 & 129 & 110 & 540 & 87.2\\ 
$^\star$2457441.64943 & -18 & 112 & 540 & 90.7\\ 
$^\star$2457441.65678 & -14 & 74 & 540 & 80.4\\ 
$^\star$2457441.66426 & 38 & 139 & 540 & 77.5\\ 
$^\star$2457441.672 & 4 & 77 & 540 & 87.6\\ 
$^\star$2457441.67973 & 4 & 90 & 540 & 81.1\\ 
$^\star$2457441.68744 & -34 & 114 & 540 & 73.7\\ 
$^\star$2457441.69538 & -151 & 143 & 540 & 67.0\\ 
$^\star$2457441.70274 & 13 & 142 & 540 & 56.7\\ 
$^\star$2457441.71029 & -5 & 96 & 540 & 60.1\\ 
$^\star$2457441.71798 & -239 & 136 & 540 & 65.5\\ 
$^\star$2457441.72596 & 175 & 98 & 540 & 74.0\\ 
$^\star$2457441.73379 & 116 & 102 & 540 & 81.1\\ 
$^\star$2457441.74174 & 326 & 101 & 540 & 78.8\\ 
$^\star$2457441.74968 & 262 & 99 & 540 & 73.0\\ 
$^\star$2457441.75734 & 78 & 128 & 540 & 70.8\\ 
$^\star$2457441.765 & 246 & 94 & 540 & 80.2\\ 
$^\star$2457441.77292 & 315 & 138 & 540 & 83.0\\ 
$^\star$2457441.78078 & 159 & 126 & 540 & 79.3\\ 
$^\star$2457441.78847 & 211 & 123 & 540 & 74.3\\ 
$^\star$2457441.79619 & 166 & 99 & 540 & 70.0\\ 
$^\star$2457441.80384 & 238 & 108 & 540 & 75.9\\ 
$^\star$2457441.81161 & 200 & 88 & 540 & 85.4\\ 
$^\star$2457441.8192 & 117 & 99 & 540 & 82.2\\ 
$^\star$2457441.82673 & -261 & 103 & 540 & 80.7\\ 
$^\star$2457441.8343 & -156 & 88 & 540 & 83.1\\ 
$^\star$2457441.84183 & -156 & 108 & 540 & 78.7\\ 
$^\star$2457441.84929 & -124 & 134 & 540 & 79.3\\ 
$^\star$2457441.85678 & -240 & 100 & 540 & 75.3\\ 
$^\star$2457441.8643 & -11 & 135 & 540 & 73.4\\ 
$^\star$2457441.87182 & -15 & 168 & 540 & 64.1\\ 
$^\star$2457441.87943 & -178 & 185 & 540 & 63.6\\ 
$^\star$2457444.65689 & 147 & 98 & 540 & 113.7\\ 
$^\star$2457444.66458 & 136 & 77 & 540 & 113.2\\ 
$^\star$2457444.6725 & 157 & 96 & 540 & 113.6\\ 
$^\star$2457444.68013 & 68 & 95 & 540 & 113.3\\ 
$^\star$2457444.68772 & 102 & 100 & 540 & 109.2\\ 
$^\star$2457444.69723 & 197 & 71 & 540 & 105.2\\ 
$^\star$2457444.70512 & 148 & 63 & 540 & 109.6\\ 
$^\star$2457444.71261 & 132 & 53 & 540 & 116.7\\ 
$^\star$2457444.72017 & 67 & 98 & 540 & 110.9\\ 
$^\star$2457444.73056 & 16 & 91 & 540 & 103.4\\ 
$^\star$2457444.73947 & 84 & 75 & 540 & 99.0\\ 
$^\star$2457444.74747 & 251 & 71 & 540 & 104.6\\ 
$^\star$2457444.75508 & 182 & 104 & 540 & 106.2\\ 
$^\star$2457444.76284 & 115 & 85 & 540 & 106.3\\ 
$^\star$2457444.77087 & 107 & 83 & 540 & 102.0\\ 
$^\star$2457444.77858 & 124 & 97 & 540 & 105.7\\ 
$^\star$2457444.78781 & 183 & 102 & 540 & 104.9\\ 
$^\star$2457444.79699 & 256 & 98 & 540 & 105.2\\ 
$^\star$2457444.80499 & 174 & 71 & 540 & 102.1\\ 
$^\star$2457444.81286 & 302 & 79 & 540 & 107.7\\ 
$^\star$2457444.82154 & 273 & 97 & 540 & 99.6\\ 
$^\star$2457444.82949 & 309 & 96 & 540 & 99.3\\ 
$^\star$2457444.83749 & 414 & 130 & 540 & 94.0\\ 
$^\star$2457444.84739 & 378 & 129 & 540 & 101.7\\ 
$^\star$2457444.85583 & 373 & 122 & 540 & 100.2\\ 
$^\star$2457444.86375 & 382 & 62 & 540 & 107.2\\ 
$^\star$2457444.8719 & 322 & 95 & 540 & 100.4\\ 
$^\star$2457444.87998 & 423 & 79 & 540 & 105.5\\ 
$^\star$2457444.88842 & 363 & 82 & 540 & 101.5\\ 
2457498.64997 & 80 & 99 & 750 & 78.8\\ 
2457499.63814 & 97 & 77 & 600 & 119.7\\ 
2457500.65083 & 49 & 96 & 900 & 107.4\\ 
2457501.62384 & -36 & 94 & 600 & 110.4\\ 
2457502.64642 & 263 & 69 & 600 & 93.7\\ 
\hline\hline
\end{tabular}
\begin{flushleft}
  \footnotesize \textbf{\textsc{NOTES}} \\
  \footnotesize $^{*}$ Exposures used to derive the Doppler tomographic transit signal, which was then used in the global EXOFAST analysis. In-transit velocities for the Rossiter-McLaughlin effect were not used in EXOFAST analysis.\\
  \footnotesize $^{a}$ \textbf{Signal to noise per resolution element of the spectrum over the Mg b line region.}
  \footnotesize $^{b}$ Template exposure defined as $0.0\,\mathrm{m\,s}^{-1}$.
  
  \end{flushleft}
  }
\end{table*}

In addition, we also observed spectroscopic transits of KELT-17b with TRES on 2016-02-23 and 2016-02-26 UT, gathering 33 and 29 sets of spectra, respectively. The exposures achieved a signal-to-noise ratio of 70--100 per resolution element over the Mg b lines, and were reduced as per \citet{Buchave:2010}. The in-transit series of spectra revealed the Doppler tomographic signal of the planet, described in Section~\ref{sec:DT}. Multi-order radial velocities were also derived for this dataset. These velocities clearly exhibit the Rossiter-McLaughlin effect \citep{Rossiter:1924,McLaughlin:1924}, are plotted in Figure~\ref{fig:K11RV}. In our global analysis with EXOFAST (Section~\ref{sec:exofast}), we model the Doppler tomographic signal, rather than the Rossiter-McLaughlin effect, to obtain the spin-orbit alignment of the system. The Doppler tomographic measurement, as presented in Section~\ref{sec:DT}, provides an accurate measurement of the spin-orbit alignment of the system. The Rossiter-McLaughlin effect, however, is modelled with an approximatation to the measured velocities due to asymmetric cross-correlation functions induced by the shadow of the planet, and are subject to modelling assumptions \citep[see discussions in e.g.][]{Boue:2013}. 



\section{Analysis and Results}

\subsection{UVW Space Motion}

We calculate the UVW space motion of KELT-17 to better understand its place in the galactic population. The proper motions, absolute radial velocities, and resulting UVW values are laid out in Table~\ref{tbl:Host_Lit_Props}. To derive the UVW space motions, we derived an absolute radial velocity measurement of KELT-17, calculated by taking the TRES Mg b echelle order absolute velocity of the template frame, subtracting a relative offset of $\gamma = 0.073\,\mathrm{km\,s}^{-1}$ from the global analysis (Section~\ref{sec:exofast}), and shifting by $-0.61\,\mathrm{km\,s}^{-1}$ to the IAU absolute velocity reference frame, which is determined by our observations of a set of IAU radial velocity reference stars. Proper motion values are taken from NOMAD \citep{Zacharias:2004}. The distance estimate is derived from a spectral fit to the spectral energy distribution (Section~\ref{sec:sed}). We also adopt the local standard of rest from \citet{2011MNRAS.412.1237C}. The resulting U, V, W space motions are $-25.6 \pm 0.9$, $3.3 \pm 0.9$, and $-0.1 \pm 1.3\,\mathrm{km\,s}^{-1}$ respectively, giving a 99.4\,\% probability that KELT-17 resides in the thin disk \citep{Bensby:2003}.

\subsection{Stellar Parameters from Spectra}
\label{sec:stellar_params}
The stellar atmospheric parameters were measured from each spectrum using the Stellar Parameter Classification (SPC) pipeline \citep{Buchave:2010}. The parameters effective temperature $T_\mathrm{eff}$, surface gravity \logg, metallicity [m/H], and projected rotational velocity $v\sin I_*$ are fitted for each TRES spectrum. SPC maximizes the cross correlation function peak of each spectrum, in the spectral order surrounding the Mg b lines, against a library of synthetic templates calculated using the \citet{Kurucz:1992} atmosphere models. The resulting stellar parameters from the first round of fitting to all exposures were $T_\mathrm{eff} = 6975 \pm 50$\,K, $\log g_*=3.08 \pm 0.10$, $\mathrm{[m/H]}= -0.10 \pm 0.08$, $v \sin I_*=49.1 \pm 0.5\, \mathrm{km\,s}^{-1}$, the uncertainties describe the expected systematic errors and scatter between exposures. The initial spectroscopic stellar parameters for rapidly rotating stars are known to be unreliable. In particular, the surface gravity \logg is difficult to determine for hot and rapidly rotating stars, and an offset will lead to systematic errors in the other atmospheric parameters. 

As such, we use the transit duration, which is directly related to the $a/R_\star$ parameter, to give a much better constraint on the stellar density $\rho_*$. Our global analysis (described in Section~\ref{sec:exofast}) simultaneously constrains the stellar parameters using the transit light curves and stellar isochrones, and yielded an updated \logg and [Fe/H].  We then re-ran SPC with the \logg fixed to that determined from the global analysis, and derived an updated set of stellar parameters of $T_\mathrm{eff}=7452 \pm 50$\,K, $\mathrm{[m/H]}=0.25 \pm 0.08$, $v\sin I_*=48.5 \pm 0.5$\,km\,s$^{-1}$. The derived $T_\mathrm{eff}$ agrees with that from the flux calibrated WiFeS low resolution spectrum (Section~\ref{sec:spec_fu}). We note the SPC-derived [m/H] is slightly different from our final metallicity value quoted in Table~\ref{tbl:KELT-17b}, since the metallicity, and \logg, are re-iterated through the global analysis, and are co-constrained by the transit-derived stellar density and the stellar isochrone models.


\subsection{SED Analysis}
\label{sec:sed}

\begin{figure}
 \centering \includegraphics[width=0.75\columnwidth, angle=90]{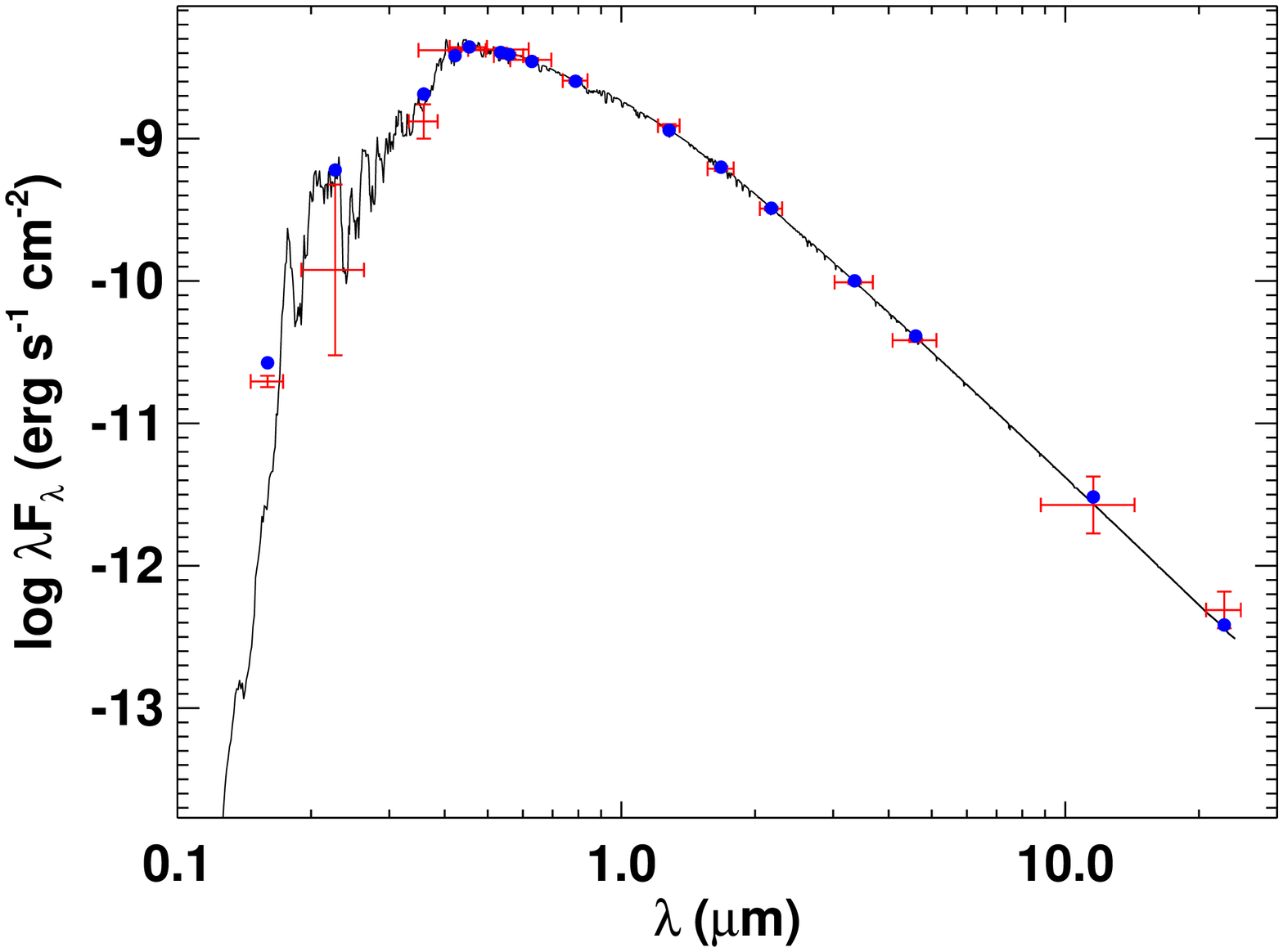}
 \caption{The SED fit for KELT-17. The red points show the photometric magnitudes and adopted uncertainties. Horizontal error bars indicate the width of the photometric band passes. The blue points show the integrated model magnitudes from the best fit NextGen synthetic spectra \citep{Hauschildt:1999}. The parameters $T_\mathrm{eff}$, $\log g_*$, [Fe/H] are allowed free in the SED fit, while the reddening $A_V$ is limited to be less than the maximum from the \citet{Schlegel:1998} dust maps.}
 \label{fig:SED_figure}
\end{figure}

We use all available broadband photometry to construct the empirical spectral energy distribution (SED) of KELT-17 (listed in Table~\ref{tbl:Host_Lit_Props}), including GALEX near-UV fluxes \citep{Bianchi:2011}, Sloan Digital Sky Survey release 7 \citep[SDSS,][]{Abazajian:2009} $u'$ band, $B_T$ and $V_T$ Tycho-2 magnitudes, All-sky Compiled Catalogue-2.5 V3 $B$ band \citep[ASCC,][]{Kharchenko:2001}, The Amateur Sky Survey Mark IV \citep[TASS,][]{Droege:2006} $V$ and $I_C$ bands, 2MASS \citep{Cutri:2003,Skrutskie:2006} $J$, $H$, $K$ bands, and WISE \citep{Cutri:2012} magnitudes (Figure~\ref{fig:SED_figure}). Only the $u'$ band magnitude is used from SDSS, as the other bands show signs of saturation. These wide band fluxes provide an independent check on the spectral classification of the host star. The SED is fitted against NextGen atmosphere models \citep{Hauschildt:1999}, with maximum reddening limited to the local value from the \citet{Schlegel:1998} dust maps. We adopt a minimum error bar of 0.03 mag if the reported error was smaller, except in the GALEX bands where  a minimum error of 0.1 mag is adopted. We also set the WISE3 error to be much larger than reported in order to account for model uncertainties at $10\,\mu m$. We derive a $T_\mathrm{eff} = 7450 \pm 150$\,K, $\log g_* = 4.0 \pm 0.5$, $\mathrm{[Fe/H]} = 0.0 \pm 0.5$, $A_V = 0.02_{-0.02}^{+0.07}$, and inferred distance of $210\pm10$\,pc, with a reduced $\chi^2$ of 4.6 from the final fit. The derived stellar parameters are in full agreement with the final SPC stellar parameters in Section~\ref{sec:stellar_params}.

\subsection{Evolutionary Analysis}

\begin{figure}
\centering
\includegraphics[width=0.75\columnwidth,angle=90]{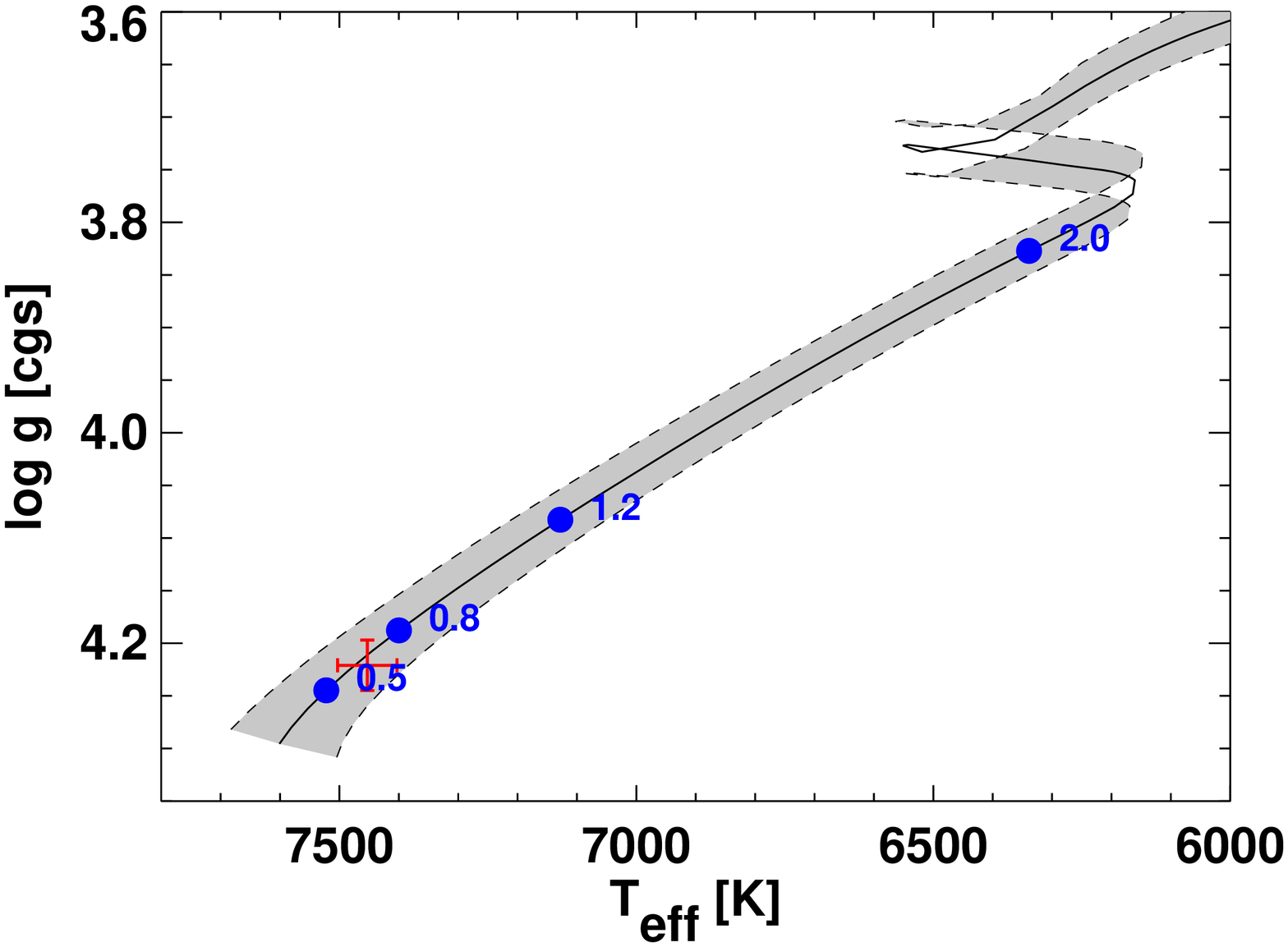}
\caption{Stellar evolutionary tracks for KELT-17. We match the $M_\star$ and [Fe/H] from the global analysis to evolutionary tracks from the YY isochrones \citep{Demarque:2004}, plotted here in terms of $T_\mathrm{eff}$ and $\log g_*$, finding an age of 0.5--0.8\,Gyr for the system. The shaded region represents the $1\sigma$ model regime, and the blue markers note the ages (in Gyr) along the best fit isochrone. The $T_\mathrm{eff}$ and $\log g_*$ of KELT-17 is marked in red.   }
\label{fig:hrd}
\end{figure}

To estimate the age of KELT-17, we match the system parameters to Yonsei-Yale (YY) evolutionary tracks \citep{Demarque:2004}, shown in Figure~\ref{fig:hrd}. Following the procedure specified in \citet{Siverd:2012} and subsequent KELT discovery papers, we adopt  $M_\star = 1.635_{-0.061}^{+0.066}\,M_\odot$ and $\mathrm{[Fe/H]} = -0.018_{-0.072}^{+0.074}$ from the global fit (Section~\ref{sec:exofast}), and match these against the YY $T_\mathrm{eff}$ -- $\log g_*$ isochrones, finding that KELT-17 is an A-star on the main-sequence with a relatively young age of 0.5--0.8\,Gyr.

\subsection{Doppler tomographic analysis}
\label{sec:DT}
\begin{figure*}
\centering
\begin{tabular}{cc}
    \includegraphics[width=8cm]{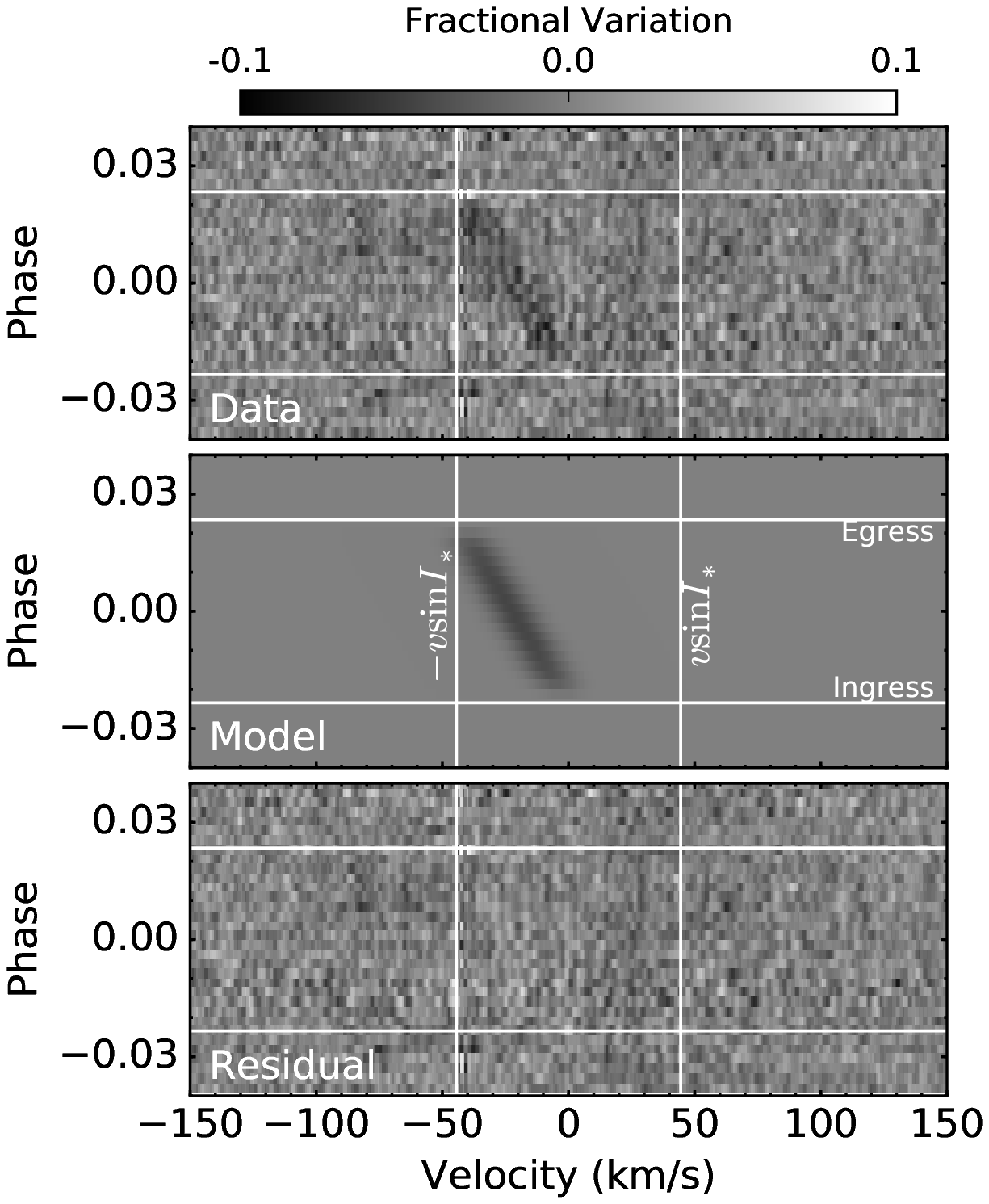} &
    \includegraphics[width=8cm]{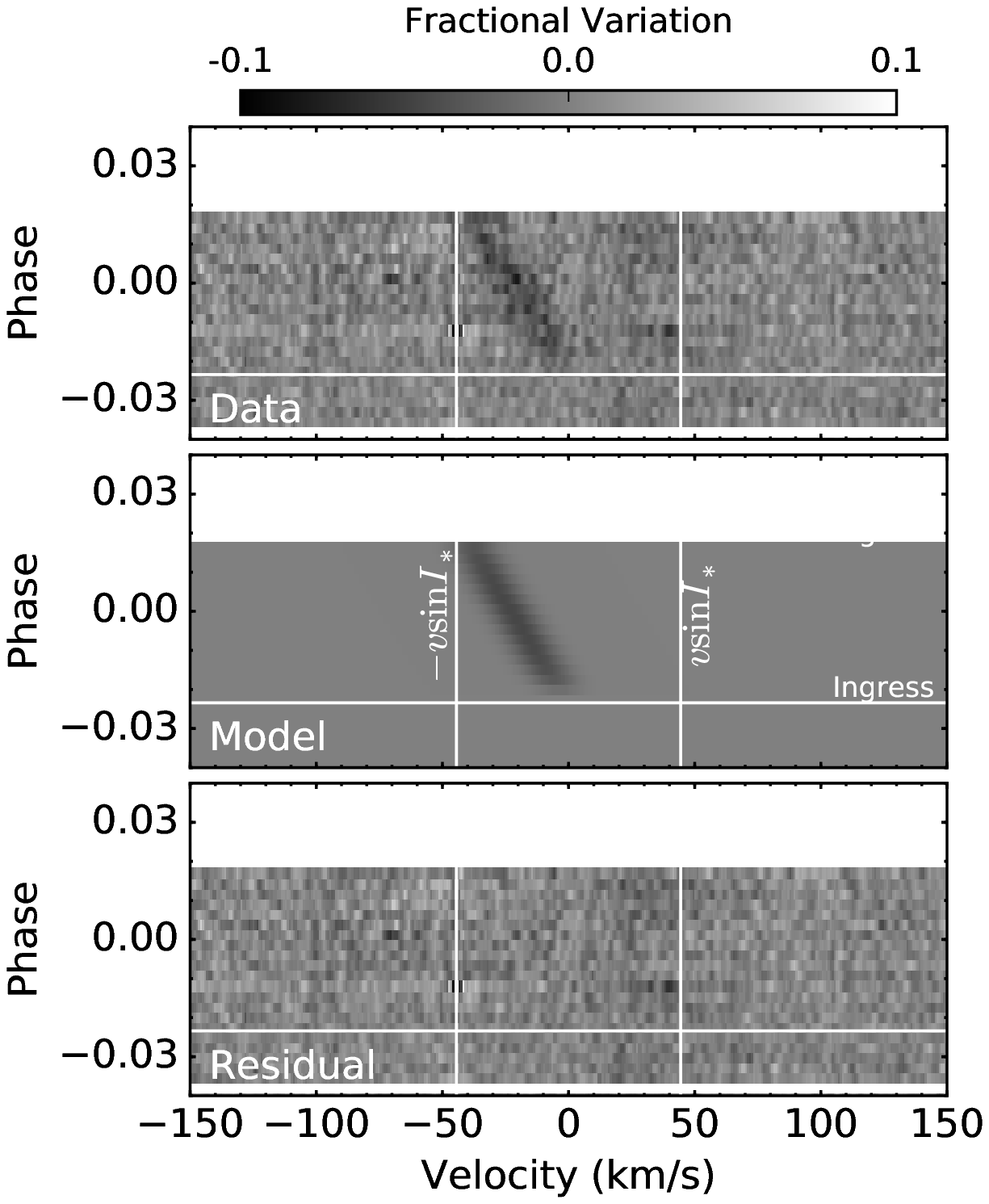} \\
\end{tabular}
\caption{\label{fig:DT} Doppler tomographic detection of the planetary transit with TRES. The transit from 2016-02-22 is plotted on the left, 2016-02-26 on the right. The top panels show the residual planetary signal after the average rotational profile is subtracted from each observation. The middle panels show the best fit model. The bottom panels show the residuals after the removal of the planetary signal. Ingress and egress are marked by the horizontal white lines. The boundaries of $-v\sin I_*$ and $v\sin I_*$ are marked by the vertical white lines.}
\end{figure*}

During a transit, the planet blocks successive regions of the star, and imprints a `shadow' on the observed spectral line profiles. For rapidly rotating stars, the line broadening profile can be derived via a least squares deconvolution of the observed spectrum against a weighted line list or an unbroadened spectral template \citep{Donati:1997,Collier:2010b}. We follow the procedure set out in \citet{Zhou:2016} to derive the broadening kernel for the set of TRES transit spectra. Each echelle order is blaze corrected by a flat lamp spectrum, and normalized by a polynomial fit to the continuum. We then stitch the spectra from every three echelle orders together, forming sections of the spectrum $\sim 200$\AA\, long. A total of 34 echelle orders were used from each observation, spanning the spectral range of 3900--6250\,\AA. For each spectral section, we generate a template using the SPECTRUM spectral synthesis program \citep{Gray:1994}, with the ATLAS9 model atmospheres \citep{Castelli:2004}. The spectral template is generated using the measured \teff, \logg, and [m/H] values from SPC, without any rotational, macroturbulence, or instrumental broadening. The broadening kernel is derived from each spectral section via the least squares deconvolution between the observed spectrum and the template \citep[as per][]{Donati:1997}. The global broadening profile of the exposure is then determined via the weighted average of the broadening kernel generated from each spectral section. Out-of-transit exposures provide an averaged broadening profile template from which the in-transit profiles are subtracted. The residuals and best fit models are shown in Figure~\ref{fig:DT}. The residuals are used in the EXOFAST global analysis in Section~\ref{sec:exofast} to help measure the spin-orbit alignment of the system, as well as co-constrain the planet transit parameters. In addition, we also derive a \vsini\ of $44.5\pm0.2\,$km\,s$^{-1}$ and a macroturbulence broadening value of $5.10\pm0.47\,\mathrm{km\,s}^{-1}$ from the deconvolved broadening kernels using the fitting technique discussed in \citet{Zhou:2016}. We also tested performing the least squares deconvolution on the SPC template spectra (calculated using \citealt{Kurucz:1992} atmosphere models), finding no measurable difference in the resulting rotational profiles.

The detection of a Doppler tomographic signal eliminates blend scenarios that can mimic the transit signal of a planetary system. The depth and width of the Doppler tomographic signal is fully consistent with the photometric transit. In blend scenarios, the Doppler tomographic signal depth will be diluted, and the width of the signal will be wider than that of a planetary signal. In particular, if the photometric transit was caused by a background eclipsing binary blended with the target star, the Doppler tomographic signal would have been undetectable.

\subsection{Transit Timing Variation Analysis}
\label{sec:TTV_analysis}
\begin{figure}[!ht]
\centering\epsfig{file=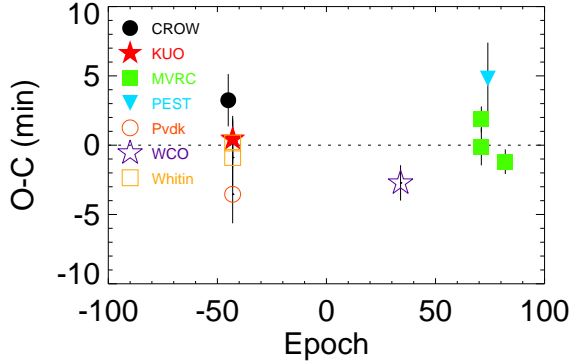,clip=,width=0.99\linewidth}
\caption{Transit time residuals for KELT-17b using the transit center times from our final global fit ephemeris. The times are listed in Table \ref{tbl:transitimes}.}
\label{fig:TTV}
\end{figure}

To determine an independent ephemeris, we perform a linear fit through the mid-transit times determined for each follow-up photometric observation (listed in Table~\ref{tbl:transitimes}). This analysis gives T$_0$ (BJD-TDB) = 2457226.142194 $\pm$ 0.00033 and a period of 3.0801718 $\pm$ 0.0000053 days, with a $\chi^2$ of 19.94 and 9 degrees of freedom. Some outliers to this fit can be seen in Figure \ref{fig:TTV}. However, the transit timing residuals are all within $2\sigma$ of a linear ephemeris, and do not show a coherent trend at levels larger than common systematic errors in transit timing \citep[e.g.][]{Carter:2009}. We carefully ensured that all follow-up observations were correctly converted to BJD$_{\rm TBD}$ \citep{Eastman:2010}. These ephemerides are then used as priors for the EXOFAST global analysis, described in Section~\ref{sec:exofast}.

\begin{table}
 \centering
 \caption{Transit times for KELT-17b.}
 \label{tbl:transitimes}
 \begin{tabular}{llllll}
    \hline
    \hline
    Epoch & $T_{C}$ 	& $\sigma_{T_{C}}$ 	& O-C &  O-C 			& Telescope \\
	  & (\bjdtdb) 	& (s) 			& (s) & ($\sigma_{T_{C}}$) 	& \\
    \hline
 -45 & 2457087.536713  & 113  &  194.59  &  1.72   &  CROW\\
 -43 & 2457093.695135  &  77   &  28.55  &  0.37   &  KUO\\
 -43 & 2457093.695102  & 101   &  25.70  &  0.25   &  KUO\\
 -43 & 2457093.692344  & 125  & -212.59  & -1.69   &  Pvdk\\
 -43 & 2457093.694912  & 106  &    9.29  &  0.09  &   Whitin\\
 -43 & 2457093.694187 &  130  &  -53.35  & -0.41  &   Whitin\\
  34 & 2457330.866145 &   76  & -163.41  & -2.13  &   WCO\\
  71 & 2457444.834316  &  80  &   -6.78  & -0.08  &   MVRC\\
  71 & 2457444.835685  &  55   & 111.50  &  1.99   &  MVRC\\
  74 & 2457454.078245  & 155  &  288.14  &  1.85  &   PEST\\
  82 & 2457478.715457 &   53  &  -71.52  & -1.34  &   MVRC\\
	  \hline
    \hline
    \hline
 \end{tabular}
\end{table}

\subsection{EXOFAST Global Fit}
\label{sec:exofast}
To provide accurate system parameters and uncertainties for the KELT-17 system, we use a modified version of the EXOFAST exoplanet fitting package \citep{Eastman:2013}, to perform a global fit of our follow-up photometric and spectroscopic observations. Here we provide an overview of the process with respect to KELT-17b; see \citet{Siverd:2012} and \citet{Kuhn:2016} for more detailed descriptions. To constrain $R_*$ and $M_*$, we adopt either the Torres relations \citep{Torres:2010} or the YY stellar evolution models \citep{Demarque:2004}. For both the Torres relations and YY Isochrones, we run a global fit constraining the eccentricity to zero. As a result of the high \vsini, causing higher than typical errors on our radial velocity measurements from TRES, we do not attempt to constrain the eccentricity. Each follow-up photometric observation (with the best determined detrending parameters from AIJ), out-of-transit radial velocity measurements from TRES, and the Doppler tomographic observations from TRES are used as inputs for the final global fits.  The results of both global fits are shown in Tables \ref{tbl:KELT-17b} and \ref{tbl:KELT-17b_part2}, and both fits are consistent with each other to within 1$\sigma$. The SPC determined \teff, [m/H], $v\sin I_*$, and line broadening due to instrumental resolution and macroturbulence $(v_\mathrm{broad})$ (and errors) from the TRES spectra, and associated $1\sigma$ uncertainties, were used as Gaussian priors during the fitting. We also adopt the period $P$ and transit epoch $T_0$ from the TTV analysis (Section~\ref{sec:TTV_analysis}) as priors in our global fitting. Allowing for TTVs decouples the transit times from the planet's orbit, adding these priors effectively encodes the information from the linear ephemeris into the global model, while still retaining the full flexibility for the transit times to vary. To simplify our interpretation, we adopt the YY circular results for the rest of this paper.

\begin{table*}
 \scriptsize
\centering
\setlength\tabcolsep{1.5pt}
\caption{Median values and 68\% confidence interval for the physical and orbital parameters of the KELT-17 system}
  \label{tbl:KELT-17b}
  \begin{tabular}{lccccc}
  \hline
  \hline
  Parameter & Description (Units) & \textbf{Adopted Value} & Value  \\
  & & \textbf{(YY circular)} & (Torres circular) \\
 \hline
Stellar Parameters & & & \\
                               ~~~$M_{*}$\dotfill &Mass (\msun)\dotfill & $1.635_{-0.061}^{+0.066}$&$1.515_{-0.071}^{+0.073}$\\
                             ~~~$R_{*}$\dotfill &Radius (\rsun)\dotfill & $1.645_{-0.055}^{+0.060}$&$1.598_{-0.054}^{+0.058}$\\
                         ~~~$L_{*}$\dotfill &Luminosity (\lsun)\dotfill & $7.51_{-0.55}^{+0.62}$&$7.07_{-0.51}^{+0.57}$\\
                             ~~~$\rho_*$\dotfill &Density (cgs)\dotfill & $0.518_{-0.042}^{+0.045}$&$0.524_{-0.044}^{+0.046}$\\
                  ~~~$\log g_*$\dotfill &Surface gravity (cgs)\dotfill & $4.220_{-0.024}^{+0.022}$&$4.211_{-0.025}^{+0.024}$\\
                  ~~~$\teff$\dotfill &Effective temperature (K)\dotfill & $7454\pm49$&$7451_{-50}^{+49}$\\
                  ~~~Age\dotfill & System Age (Gyr)\dotfill & 0.5--0.8 \\
                                 ~~~$\feh$\dotfill &Metallicity\dotfill & $-0.018_{-0.072}^{+0.074}$&$-0.274_{-0.072}^{+0.11}$\\
             ~~~$v\sin{I_*}$\dotfill &Rotational velocity $(\mathrm{m\,s}^{-1})$\dotfill & $44200_{-1300}^{+1500}$&$44100_{-1300}^{+1500}$\\
           ~~~$\lambda$\dotfill & Projected spin-orbit alignment (degrees)\dotfill & $-115.9\pm4.1$&$-115.5_{-4.2}^{+4.1}$\\
           ~~~$I_*$ $^a$\dotfill & Line-of-sight stellar inclination (degrees)\dotfill & $94_{-10}^{+9}$ & \\
           ~~~$\phi$ $^a$\dotfill & True obliquity (degrees)\dotfill & $116\pm4$ &\\
         ~~~$ v_\mathrm{broad} $\dotfill & non-rotational line broadening $(\mathrm{m\,s}^{-1})$\dotfill & $5100\pm470$&$5080\pm470$\\
\hline
 Planet Parameters & & & \\
                                  ~~~$P$\dotfill &Period (days)\dotfill & $3.0801716_{-0.0000052}^{+0.0000053}$&$3.0801718_{-0.0000038}^{+0.0000037}$\\
                           ~~~$a$\dotfill &Semi-major axis (AU)\dotfill & $0.04881_{-0.00061}^{+0.00065}$&$0.04759\pm0.00075$\\
                                 ~~~$M_{P}$\dotfill &Mass (\mj)\dotfill & $1.31_{-0.29}^{+0.28}$&$1.25\pm0.27$\\
                               ~~~$R_{P}$\dotfill &Radius (\rj)\dotfill & $1.525_{-0.060}^{+0.065}$&$1.478_{-0.058}^{+0.062}$\\
                           ~~~$\rho_{P}$\dotfill &Density (cgs)\dotfill & $0.46_{-0.11}^{+0.12}$&$0.48_{-0.11}^{+0.12}$\\
                      ~~~$\log{g_{P}}$\dotfill &Surface gravity (cgs)\dotfill & $3.144_{-0.11}^{+0.090}$&$3.149_{-0.11}^{+0.089}$\\
               ~~~$T_{eq}$\dotfill &Equilibrium temperature (K)\dotfill & $2087_{-33}^{+32}$&$2082_{-32}^{+34}$\\
                           ~~~$\Theta$\dotfill &Safronov number\dotfill & $0.051\pm0.011$&$0.053\pm0.011$\\
                   ~~~$\fave$\dotfill &Incident flux (\fluxcgs)\dotfill & $4.31_{-0.26}^{+0.27}$&$4.26_{-0.25}^{+0.28}$\\
\hline
 Radial Velocity Parameters & & & \\
       ~~~$T_C$\dotfill &Time of inferior conjunction (\bjdtdb)\dotfill & $2457287.74564\pm0.00030$&$2457287.74565\pm0.00021$\\
                        ~~~$K$\dotfill &RV semi-amplitude $(\mathrm{m\,s}^{-1})$\dotfill & $131_{-29}^{+28}$&$131\pm28$\\
                    ~~~$M_P\sin{i}$\dotfill &Minimum mass (\mj)\dotfill & $1.31_{-0.29}^{+0.28}$&$1.24\pm0.27$\\
                           ~~~$M_{P}/M_{*}$\dotfill &Mass ratio\dotfill & $0.00077\pm0.00017$&$0.00079\pm0.00017$\\
                       ~~~$u$\dotfill &RM linear limb darkening\dotfill & $0.5383_{-0.0020}^{+0.0028}$&$0.5437_{-0.0051}^{+0.014}$\\
                                ~~~$\gamma_{TRES}$ \dotfill & Offset for TRES relative velocities $(\mathrm{m\,s}^{-1})$\dotfill & $73\pm24$&$74\pm24$\\
\hline
Linear Ephemeris & &  &            \\
from Follow-up & &  &            \\
Transits: & &  &            \\
                                  ~~~$P_{Trans}$\dotfill &Period (days)\dotfill & $3.0801718\pm 0.0000053$&---\\
       ~~~$T_0$\dotfill &Linear ephemeris from transits (\bjdtdb)\dotfill & $2457226.142194\pm0.00033$ &---\\
 \hline
 \hline
 \hline
 \end{tabular}
\begin{flushleft}
 \footnotesize \textbf{\textsc{NOTES}} \\
  \vspace{.1in}
  \footnotesize $^{a}$ From the independent differential rotation analysis described in Section~\ref{sec:diffrot}.
 \end{flushleft}
\end{table*}

\begin{table*}
\scriptsize
 \centering
\setlength\tabcolsep{1.5pt}
\caption{Median values and 68\% confidence intervals for the physical and orbital parameters for the KELT-17 System}
  \label{tbl:KELT-17b_part2}
  \begin{tabular}{lccccc}
  \hline
  \hline
  Parameter & Description (Units) & \textbf{Adopted Value} & Value \\
  & & \textbf{(YY circular)} & (Torres circular) \\
 \hline
 \hline
 Primary Transit & & & \\
~~~$R_{P}/R_{*}$\dotfill &Radius of the planet in stellar radii\dotfill & $0.09526_{-0.00085}^{+0.00088}$&$0.09509\pm0.00086$\\
           ~~~$a/R_*$\dotfill &Semi-major axis in stellar radii\dotfill & $6.38\pm0.18$&$6.40\pm0.18$\\
                          ~~~$i$\dotfill &Inclination (degrees)\dotfill & $84.87_{-0.43}^{+0.45}$&$84.93\pm0.45$\\
                               ~~~$b$\dotfill &Impact parameter\dotfill & $0.570_{-0.035}^{+0.031}$&$0.566_{-0.035}^{+0.033}$\\
                             ~~~$\delta$\dotfill &Transit depth\dotfill & $0.00907_{-0.00016}^{+0.00017}$&$0.00904\pm0.00016$\\
                    ~~~$T_{FWHM}$\dotfill &FWHM duration (days)\dotfill & $0.12674_{-0.00067}^{+0.00068}$&$0.12667_{-0.00064}^{+0.00065}$\\
              ~~~$\tau$\dotfill &Ingress/egress duration (days)\dotfill & $0.0181_{-0.0011}^{+0.0012}$&$0.0179_{-0.0011}^{+0.0012}$\\
                     ~~~$T_{14}$\dotfill &Total duration (days)\dotfill & $0.1448_{-0.0013}^{+0.0014}$&$0.1446_{-0.0013}^{+0.0014}$\\
   ~~~$P_{T}$\dotfill &A priori non-grazing transit probability\dotfill & $0.1418_{-0.0038}^{+0.0039}$&$0.1413_{-0.0038}^{+0.0041}$\\
             ~~~$P_{T,G}$\dotfill &A priori transit probability\dotfill & $0.1717_{-0.0048}^{+0.0050}$&$0.1710_{-0.0048}^{+0.0051}$\\
                     ~~~$u_{1B}$\dotfill &Linear Limb-darkening\dotfill & $0.3713_{-0.0053}^{+0.0064}$&$0.3795_{-0.0098}^{+0.020}$\\
                  ~~~$u_{2B}$\dotfill &Quadratic Limb-darkening\dotfill & $0.3509_{-0.0040}^{+0.0034}$&$0.3462_{-0.0100}^{+0.0062}$\\
                     ~~~$u_{1I}$\dotfill &Linear Limb-darkening\dotfill & $0.1337_{-0.0032}^{+0.0047}$&$0.1504_{-0.0075}^{+0.023}$\\
                  ~~~$u_{2I}$\dotfill &Quadratic Limb-darkening\dotfill & $0.3266_{-0.0027}^{+0.0024}$&$0.3122_{-0.019}^{+0.0056}$\\
                ~~~$u_{1Sloang}$\dotfill &Linear Limb-darkening\dotfill & $0.3418_{-0.0040}^{+0.0054}$&$0.3527_{-0.0087}^{+0.022}$\\
             ~~~$u_{2Sloang}$\dotfill &Quadratic Limb-darkening\dotfill & $0.3480_{-0.0033}^{+0.0024}$&$0.3408_{-0.012}^{+0.0055}$\\
                ~~~$u_{1Sloani}$\dotfill &Linear Limb-darkening\dotfill & $0.1511_{-0.0032}^{+0.0047}$&$0.1680_{-0.0083}^{+0.025}$\\
             ~~~$u_{2Sloani}$\dotfill &Quadratic Limb-darkening\dotfill & $0.3315\pm0.0025$&$0.3173_{-0.020}^{+0.0059}$\\
                ~~~$u_{1Sloanr}$\dotfill &Linear Limb-darkening\dotfill & $0.2179_{-0.0029}^{+0.0046}$&$0.2264_{-0.0096}^{+0.032}$\\
             ~~~$u_{2Sloanr}$\dotfill &Quadratic Limb-darkening\dotfill & $0.3501_{-0.0022}^{+0.0019}$&$0.3426_{-0.023}^{+0.0067}$\\
                ~~~$u_{1Sloanz}$\dotfill &Linear Limb-darkening\dotfill & $0.1023_{-0.0033}^{+0.0047}$&$0.1175_{-0.0058}^{+0.018}$\\
             ~~~$u_{2Sloanz}$\dotfill &Quadratic Limb-darkening\dotfill & $0.3194_{-0.0029}^{+0.0020}$&$0.3060_{-0.015}^{+0.0041}$\\
                     ~~~$u_{1V}$\dotfill &Linear Limb-darkening\dotfill & $0.2795_{-0.0028}^{+0.0045}$&$0.2911_{-0.0091}^{+0.028}$\\
                  ~~~$u_{2V}$\dotfill &Quadratic Limb-darkening\dotfill & $0.3467_{-0.0025}^{+0.0016}$&$0.3384_{-0.018}^{+0.0054}$\\
\hline
Secondary Eclipse & & & \\
                  ~~~$T_{S}$\dotfill &Time of eclipse (\bjdtdb)\dotfill & $2457286.20555\pm0.00030$&$2457286.20556\pm0.00021$\\
     \hline
 \hline
\end{tabular}
\end{table*}

Since the follow-up light curves were obtained in multiple photometric bands, we can also search for signs of wavelength -- transit depth dependencies. Large color-based transit depth dependencies can indicate the transit is actually caused by a stellar eclipsing binary system, while low level trends can reveal Rayleigh scatter signatures in the planetary atmospheres. The follow-up data includes observations performed in the $B$, $g'$, $V$, $R$, $i'$, and $z'$ bands. We allowed the transit model for each band to have independent $R_p/R_\star$ values, all other transit parameters are shared in the joint fitting. The limb darkening coefficients are fixed to those interpolated from \citet{Claret:2011}. We find no color-transit depth dependencies, with all derived $R_p/R_\star$ values agreeing to within $1\sigma$.

\subsection{Constraining differential rotation via Doppler tomography}
\label{sec:diffrot}

Planets in strongly misaligned orbits can allow us to map the surface features on the host star. For example, spot-crossings during the transits of the polar orbit planet HAT-P-11b were used to construct a `butterfly-diagram' for the spot evolution of the K-star \citep{Sanchis:2011}. Similarly, the Doppler tomographic shadow of a spin-orbit misaligned planetary transit can help map the projected spin velocity of the stellar surface underneath. In particular, this allows us to directly measure the stellar spin rate as a function of latitude \citep{Gaudi:2007,Cegla:2016}, thereby constraining the latitudinal differential rotation rate of the host star. 

On the sun, differential rotation is thought to partially drive the solar dynamo \citep[e.g.][]{Dikpati:1999}, integral to the development of sun-spots and the 11-year solar activity cycle. Non-rigid rotation has been inferred for other stars by monitoring for long-term activity cycles in their spot modulated light curves \citep[e.g.][]{Walkowicz:2012,Reinhold:2013}, via Doppler tomographic maps of spotted, active stars \citep[e.g.][]{Donati:1997b,Barnes:2000}, or via modeling of the rotational broadening kernel \citep[e.g.][]{Reiners:2002,Reiners:2003,Reiners:2003b}. Planetary transits also offer a method of directly detecting differential rotation for favorable systems.

Constraining differential rotation with planetary systems is also particularly interesting in the context of spin-orbit misaligned hot-Jupiter systems, like KELT-17b. A large fraction of hot-Jupiter systems are reported to be spin-orbit misaligned, with the G, K stars hosting a larger fraction of aligned systems than stars of earlier spectral types \citep[e.g.][]{Winn:2010,Albrecht:2012}. One idea is that internal gravity waves generated at the radiative envelope -- convective core boundaries of hot stars can modify the apparent rotation of these stars at short timescales, in which case both radial and latitudinal differential rotation are expected \citep{Rogers:2012,Rogers:2013b}. Direct measurements of host star differential rotation of a spin-orbit misaligned system is key to testing this idea.

\begin{figure*}
\centering
\includegraphics[width=14cm]{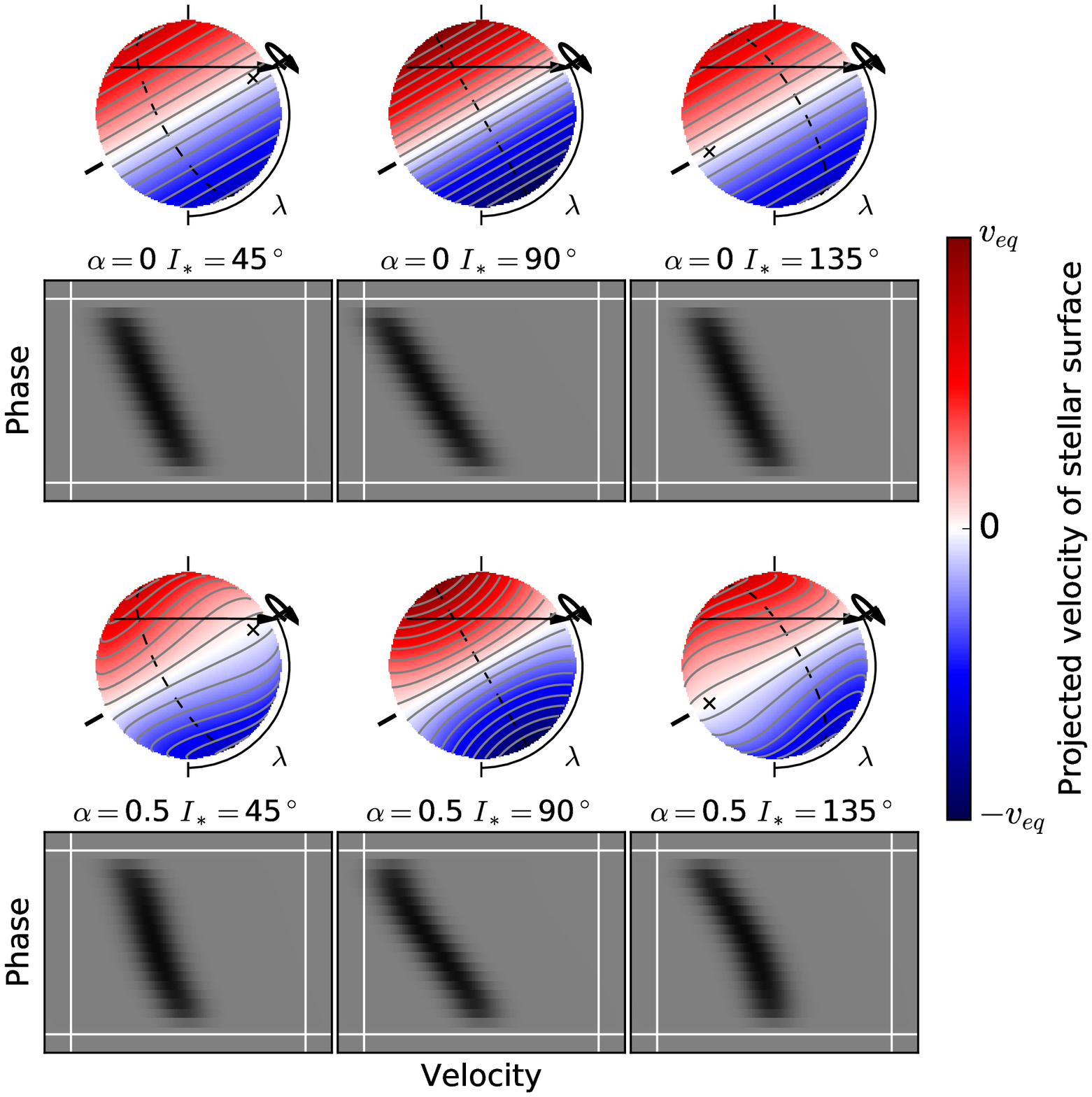}

\caption{\label{fig:dfmodels}A spin-orbit misaligned planet that transits across different latitudes of the host star can be used to probe for latitudinal differential rotation on the host star. We simulate the Doppler tomographic signal of KELT-17b for host stars exhibiting different levels of differential rotation $(\alpha)$, at various line-of-sight inclinations $(I_*)$. Differential rotation manifests in the Doppler tomographic signal as a curved trail. The simulated planet transit parameters are the same as those derived for KELT-17b in Table~\ref{tbl:KELT-17b}, including the projected spin-orbit alignment value $\lambda$. The orbit normal are marked by the vertical dashes above and below the stellar disk. Black arrows represent the path of planet. The contour lines represent equal projected velocity regions on the stellar surface. The crosses mark the rotational poles of the stars, and the thick diagonal lines mark the rotation axis. The dashed line marks the equator. The angle $\lambda$ is also labelled. On the Doppler tomographic diagrams, white lines represent ingress and egress and $\pm v \sin I_*$ as per Figure~\ref{fig:DT}. In the individual panels, we demonstrate the effect of two $\alpha$ parameters -- solid body $(\alpha=0)$ and large levels of differential rotation $(\alpha=0.5)$, at three values of $I_*$, to exaggerate the effect of differential rotation for the reader. }
\end{figure*}

We attempt to fit for any differential rotation in KELT-17 via an analysis that is independent, and simplified, from that of the EXOFAST fitting described in Section~\ref{sec:exofast}. Following \citet{Cegla:2016}, we model the stellar rotation at a given position, $v_\mathrm{stel}$ as  
\begin{equation}
    v_\mathrm{stel} = x v_\mathrm{eq}\sin I_* ( 1 - \alpha y^2)\, 
\end{equation}
where $x$ and $y$ are the projected position coordinates on the stellar disk with respect to the stellar spin axis and equator, $v_\mathrm{eq}$ is the equatorial rotation speed, and $I_*$ is the line-of-sight inclination of the stellar spin-axis. The coefficient $\alpha$ describes the rate of differential rotation, where a rigid body has $\alpha=0$, while the solar differential rotation is described by $\alpha_\odot = 0.2$.  The effect of various levels of differential rotation $\alpha$ and inclination $I_*$ for the KELT-17 system on the observed Doppler tomographic maps are illustrated in Figure~\ref{fig:dfmodels}. 

We incorporate the differential rotation model in an independent global fit of the follow-up light curves and Doppler tomographic datasets, fitting for the differential rotation parameters $\alpha$ and $I_*$, transit parameters period $P$, transit time $T_0$, radius ratio $R_p/R_\star$, normalized semi-major axis $a/R_\star$, transit chord inclination $i$, projected spin-orbit angle $\lambda$, projected rotation velocity $v\sin I_*$, stellar parameters \teff, \logg, and first order light curve detrending coefficients for each of the instrumental trends set out in Table~\ref{tbl:detrending_parameters}. The fitting procedure is largely described in \citet{Zhou:2016}, and is performed via a MCMC analysis with the \emph{emcee} affine invariant ensemble sampler \citep{ForemanMackey:2012}.

In \citet{Zhou:2016}, the tomographic signal of the planet is approximated by a Gaussian of width $v\sin I_* \times R_p/R_\star$, sufficient for the standard modeling of the planetary Doppler tomographic signal. However, a true model of the planetary shadow profile needs to account for asymmetries during ingress and egress, as well as the uneven limb darkening in the projected stellar surface under the planet. This is especially important when fitting for differential rotation, which relies on accurate centroids of the planetary shadow at each time step. Therefore, we model the planetary shadow via a numerical integration of the projected stellar surface underneath the planet, accounting for limb darkening, differential rotation, and instrumental broadening. 

\begin{figure*}
\centering
\includegraphics[width=17cm]{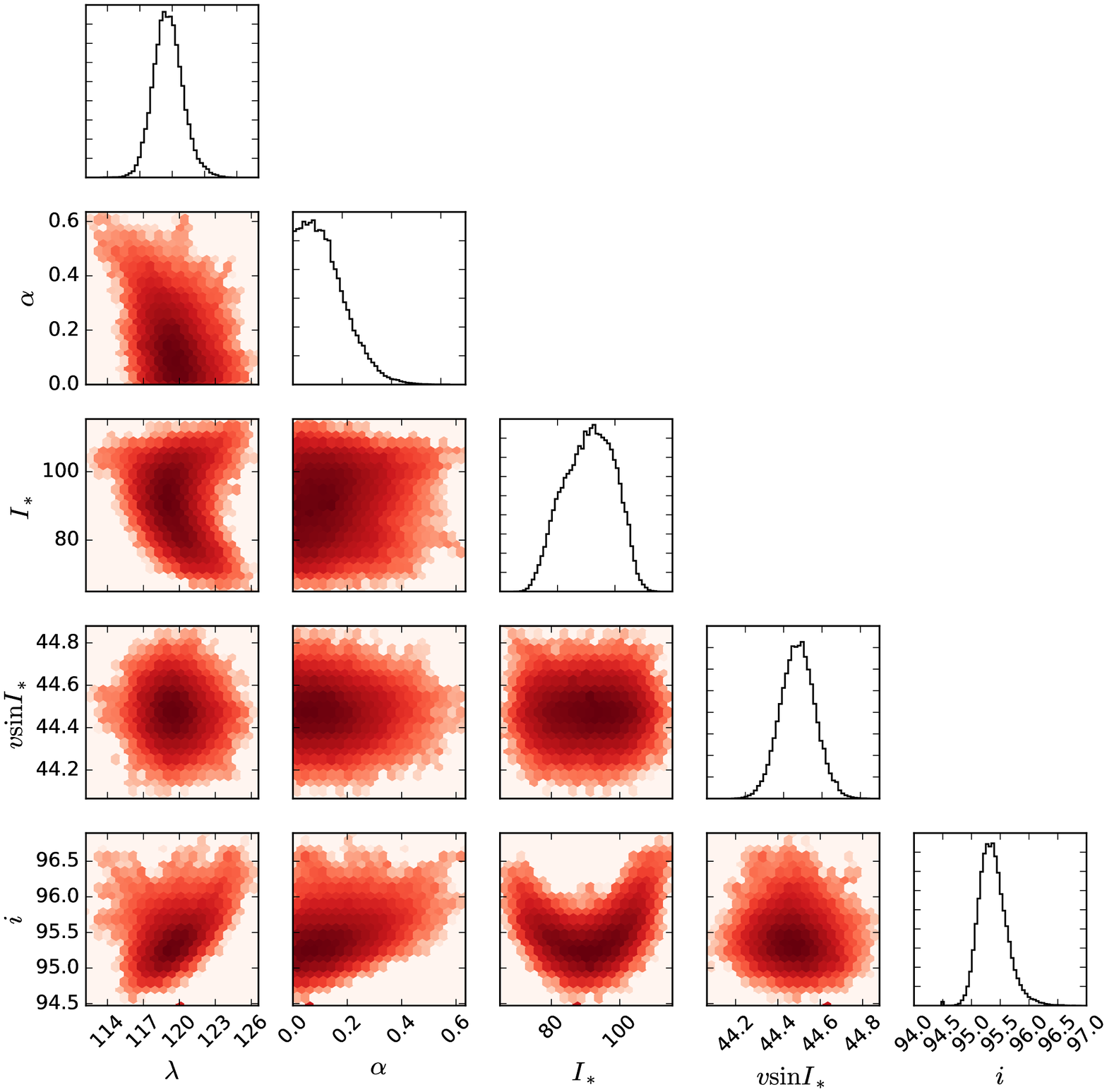}

\caption{\label{fig:dfcorner}Posterior probability distributions for the differential rotation modeling, showing constraints on the parameters $\alpha$ and $I_*$, as well as selected system parameters such as transit chord inclination ($i$), projected spin-orbit alignment $\lambda$, and projected rotational velocity $v \sin I_*$. We can constrain $\alpha<0.30$ at $2\sigma$ significance.}
\end{figure*}

Figure~\ref{fig:dfcorner} shows the posterior distribution of selected parameters from our MCMC analysis. The differential rotation coefficient can be constrained to be $\alpha < 0.15$ at $1\sigma$ ($\alpha < 0.30$ at $2\sigma$), consistent with rigid body rotation, but also consistent with solar differential rotation. The line-of-sight inclination is constrained to $I_* = 94 _{-10}^{+9}$ deg. The line-of-sight inclination $I_*$ and the projected spin-orbit angle $\lambda$ can be combined to calculate the true spin-orbit angle $\phi$:
\begin{equation}
    \cos \phi = \cos I_* \cos i + \sin I_* \sin i \cos \lambda \, ,
\end{equation}
giving a true spin-orbit angle of $\phi = 116\pm4\,^\circ$ for the system. 

\citet{AmmlervonEiff:2012} found the fraction of differential rotators decreases with increasing $T_\mathrm{eff}$ and $v\sin I_*$ for A-F stars. KELT-17 is a late A-star with no evidence for differential rotation, consistent with this trend. In contrast, Doppler imaging of spots on rapid rotators from \citet{Barnes:2005}, \citet{CollierCameron:2007}, and models from \citet{Kuker:2005} find the level of differential rotation to increase with temperature for late F to M stars. Direct detections of differential rotation via spin-orbit misaligned planets for a range of stars can be a way to provide clear benchmarks to re-examine these previous studies. Under the framework of \citet{Rogers:2012}, the lack of significant differential rotation suggests there is no evidence that the star is currently undergoing spin axis evolution.

\section{Conclusions}

We report the discovery of KELT-17b, a hot-Jupiter around an A-star discovered by the joint KELT-North and KELT-South survey. KELT-17b is only the fourth hot-Jupiter found transiting an A-star, after WASP-33b, KOI-13b, and HAT-P-57b. The host star is also amongst the most rapidly rotating known transit-planet-hosting stars, with a $v\sin I_*$ of $44.2_{-1.3}^{+1.5}\,\mathrm{km\,s}^{-1}$, only WASP-33 \citep{Collier:2010b}, KELT-1 \citep{Siverd:2012}, KOI-12 \citep{Bourrier:2015}, KOI-13 \citep{Szabo:2011}, KELT-7 \citep{Bieryla:2015}, and HAT-P-57 \citep{Hartman:2015} exhibit faster spin rates. KELT-17, with a mass of $1.635_{-0.061}^{+0.066}\, M_\odot$, is amongst the most massive (hottest) 3\% (0.5\%) of known planet hosts\footnote{NASA exoplanet archive \url{http://exoplanetarchive.ipac.caltech.edu/}}. Blend scenarios for KELT-17b are eliminated by the detection of the Doppler tomographic signal, from which we also measured a projected spin-orbit angle of $\lambda = -115.9\pm4.1^\circ$ for the system. With a mass of $1.31_{-0.29}^{+0.28}\,M_J$, and radius of $1.525_{-0.060}^{+0.065}\,R_J$, KELT-17b is inflated compared to standard evolution models. It receives an incident flux of $4\times 10^9\,\mathrm{erg\,s}^{-1}\mathrm{cm}^{-2}$, larger than the empirical threshold of $2\times10^8\,\mathrm{erg\,s}^{-1}\mathrm{cm}^{-2}$ where radius inflation is observed for the hot-Jupiter sample \citep{Demory:2011}.

KELT-17b is one of 26 known transiting hot-Jupiters around a host star hotter than 6250\,K, of which 70\% are spin-orbit misaligned ($|\lambda| > 10^\circ$). In fact, all four hot-Jupiters around A-stars are in severely misaligned orientations\footnote{Multiple $\lambda$ solutions are allowed for HAT-P-57b \citep{Hartman:2015}}. The spin-orbit synchronization timescale for the KELT-17 system is $\sim 10^{11}$ yrs \citep[using Equation 3,][]{Hansen:2012}, so the current system misalignment is unlikely to have been modified by star-planet tidal interactions, and will be stable for the duration of the main-sequence lifetime of KELT-17. KELT-17b is super-synchronous: the host star has a maximum spin period of 1.7 days, while the planet orbital period is $\sim 3.08$ days, as with a number of other systems around F-A stars (CoRoT-3b, CoRoT-11b, HAT-P-56b, KELT-7b, KOI-13b, WASP-7b, WASP-8b, WASP-33b, WASP-38b). In contrast, no Kepler candidates, which are largely around cooler host stars, are found in super-synchronized orbits \citep{Walkowicz:2013}. While the synchronization timescale is long, the orbit circularization timescale should be only $10^7$ years, assuming $Q_\mathrm{planet} = 10^5$, and adopting the circularization timescale from \citet{Goldreich:1966}, so we expect the orbit of the planet to be circular in the present day.

The spin-orbit misaligned orientation of KELT-17b means the planet crosses a wide-range of stellar latitudes during the transit, which allowed us to constrain the latitudinal differential rotation of the star. As a result, we find KELT-17 to be consistent with both rigid body rotation and solar-levels of differential rotation  ($\alpha<0.30$ at $2\sigma$). An equivalent technique was applied to the transits of HD 189733b \citep{Cegla:2016}, a significantly more difficult case given the near-aligned geometry of the planet $(\lambda = -0.4\pm 0.2^\circ)$ and the low rotation rate of the star $(v\sin I_* = 3.25\pm0.02\,\mathrm{km\,s}^{-1})$. Nevertheless, they were able to rule out rigid-body rotation for the host star. Future Doppler tomographic follow-up of KELT-17 can further refine the differential rotation constraints on the star, and search for nodal precession of the planet's orbit \citep[e.g.][]{Johnson:2015}.

\acknowledgements
The authors thank the referee for their insightful comments. K.K. McLeod acknowledges the Theodore Dunham, Jr. Grant of the Fund for Astronomical Research for the purchase of the SDSS filters used at Whitin Observatory. B.J.F. notes that this material is based upon work supported by the National Science Foundation Graduate Research Fellowship under grant No. 2014184874. Work by B.S.G. and D.J.S was partially supported by NSF CAREER Grant AST-1056524. Any opinion, findings, and conclusions or recommendations expressed in this material are those of the authors(s) and do not necessarily reflect the views of the National Science Foundation.

\bibliographystyle{apj}

\bibliography{KELT-17b}

\end{document}